\documentclass[fleqn,usenatbib]{mnras}

\usepackage{newtxtext,newtxmath}
\usepackage[T1]{fontenc}

\DeclareRobustCommand{\VAN}[3]{#2}
\let\VANthebibliography\thebibliography
\def\thebibliography{\DeclareRobustCommand{\VAN}[3]{##3}\VANthebibliography}

\usepackage{graphicx}	% Including figure files
\usepackage{amsmath}	% Advanced maths commands
\usepackage{pdflscape}
\usepackage{ulem}

\newcommand{\fracb}[2]{\left(\frac{#1}{#2}\right)}

\usepackage[dvipsnames]{xcolor}
\definecolor{blazeorange}{rgb}{1.0, 0.4, 0.0}
\definecolor{seagreen}{rgb}{0.18, 0.55, 0.34}
\definecolor{rufous}{rgb}{0.66, 0.11, 0.03}
\definecolor{royalfuchsia}{rgb}{0.79, 0.17, 0.57}
\definecolor{scarlet}{rgb}{1.0, 0.13, 0.0}
\definecolor{royalpurple}{rgb}{0.47, 0.32, 0.66}

\newcommand\confirm[1]{#1}

\title[GRB Polarimetry with POLAR-2]{Prospects of Prompt Gamma-Ray Burst Polarimetry with POLAR-2}

\author[Gill et al. 2025]{
Ramandeep Gill,$^{1,2}$\thanks{E-mail: rsgill.rg@gmail.com (RG)}
Jiang He,$^{3}$\thanks{E-mail: hejiang@ihep.ac.cn (JH)}
Jonathan Granot,$^{2,4}$\thanks{E-mail: granot.j@gmail.com (JG)}
Jian-Chao Sun,$^{3}$\thanks{E-mail: sunjc@ihep.ac.cn (JS)}
Shuang-Nan Zhang,$^{3,5}$
\newauthor{Yuan-Hao Wang,$^{3}$
Johannes Hulsman,$^{6}$
Nicolas Produit,$^{7}$
Shao-Lin Xiong$^{3}$
}
\\
$^{1}$Instituto de Radioastronom\'ia y Astrof\'isica, Universidad Nacional Aut\'onoma de M\'exico, Antigua Carretera a P\'atzcuaro $\#$ 8701,  Ex-Hda. San Jos\'e de la \\ Huerta, Morelia, Michoac\'an, C.P. 58089, M\'exico\\
$^{2}$Astrophysics Research Center of the Open university (ARCO), The Open University of Israel, P.O Box 808, Ra'anana 43537, Israel\\
$^{3}$State Key Laboratory of Particle Astrophysics, Institute of High Energy Physics, Chinese Academy of Sciences, Beijing 100049, China\\
$^{4}$Department of Natural Sciences, The Open University of Israel, P.O Box 808, Ra'anana 43537, Israel\\
$^{5}$University of Chinese Academy of Sciences, Chinese Academy of Sciences, Yuquan Road, Shijingshan District, Beijing, 100049, China\\
$^{6}$DPNC, University of Geneva, 24 Quai Ernest-Ansermet, Geneva, CH-1205, Switzerland\\
$^{7}$Geneva Observatory, ISDC, University of Geneva, 16, Chemin
d’Ecogia, Versoix, CH-1290, Geneva, Switzerland
}

\date{Accepted XXX. Received YYY; in original form ZZZ}

\pubyear{2023}

\begin{document}
\label{firstpage}
\pagerange{\pageref{firstpage}--\pageref{lastpage}}
\maketitle

\begin{abstract}
The dominant radiation mechanism that powers the prompt $\gamma$-ray emission in gamma-ray bursts (GRBs) remains poorly understood. High quality, time- and energy-resolved linear polarization measurements of prompt $\gamma$-ray photons can distinguish between synchrotron and inverse-Compton processes and provide crucial constraints on the outflow properties. This will be achieved by POLAR-2 that is proposed as a dedicated GRB polarimeter and successor to POLAR. The High-energy Polarimetry Detector (HPD) is one of the three instruments of POLAR-2 that features significantly improved sensitivity in the $(40-1000)$\,keV energy range and a detection area four times larger than that of POLAR. Here we demonstrate the capabilities of the HPD to constrain key physical model parameters by creating and fitting to synthetic sources using a time-resolved spectro-polarimetric theoretical model of prompt GRB emission. The time-resolved spectral and polarization fits are performed using a novel technique featuring maximum likelihood over an unbinned (in time and energy) list of detected events. The constrained model parameters directly relate to the underlying source physics that would reveal an accelerating, coasting or decelerating emission region. For a pulse fluence of $\mathcal{F}=10^{-5}\mathcal{F}_{-5}\,{\rm erg\,cm^{-2}}$ \confirm{and higher} we can constrain the time-integrated polarization degree to an absolute accuracy ($1\,\sigma$) of about $\confirm{2.2}\mathcal{F}_{-5}^{\,-1/2}$ per cent, as long as source photons dominate over the background. In bright GRBs, such unprecedented accuracy at these energies will allow to distinguish between different models for the prompt GRB emission mechanism and constrain the magnetic field geometry, jet angular structure and outflow composition.
\end{abstract}


%%%%%%%%%%%%%%%%%%%%%%%%%%%%%%%%%%%%%%%%%%%%%%%%%%

%%%%%%%%%%%%%%%%% BODY OF PAPER %%%%%%%%%%%%%%%%%%

%%%%%%%%%%%%%%%%%%%%%%%%%%%%%%%%%%%%%%%%%%%%%%%%%%%%%%%
\section{Introduction}
%%%%%%%%%%%%%%%%%%%%%%%%%%%%%%%%%%%%%%%%%%%%%%%%%%%%%%%
Gamma-ray bursts (GRBs) are the most luminous high-energy transients in our Universe, powered by ultra-relativistic jets with bulk Lorentz factors (LFs) $\Gamma\gtrsim100$ \citep[see, e.g.,][for comprehensive reviews]{Piran-04,Meszaros-06,Kumar-Zhang-15}. Internal dissipation within the jets converts part of the kinetic and/or magnetic field energy in the ejecta to internal energy accelerating particles that produce extremely high peak $\gamma$-ray luminosities of $L_{\gamma,\rm iso}\sim10^{51-54}\,{\rm erg\,s}^{-1}$ (isotropic-equivalent), making GRBs visible up to cosmological distances. The initial burst of $\gamma$-rays, dubbed the \textit{prompt} emission, has a non-thermal spectrum and in the majority of cases it is described by an empirical Band-function \citep{Band+93} that features a smoothly broken power-law. Despite efforts over the past several decades to understand the radiation mechanism that powers the prompt GRB, it remains one of the most important open questions in GRB physics. Two radiation mechanisms, namely synchrotron radiation from shock accelerated relativistic electrons with a power-law energy distribution \citep{Sari-Piran-97,Daigne-Mochkovitch-98} and Comptonized emission of softer seed thermal photons by mildly relativistic electrons \citep[e.g.][]{Thompson-94,Ghisellini-Celotti-99,Meszaros-Rees-00,Rees-Meszaros-05,Giannios-06,Peer+06,Beloborodov-10,Thompson-Gill-14,Gill-Thompson-14}, have emerged as the most favoured explanations for the prompt spectrum.

%%%%% FIGURE %%%%%%%%%
\begin{figure*}
\centering
\includegraphics[width=0.6\textwidth]{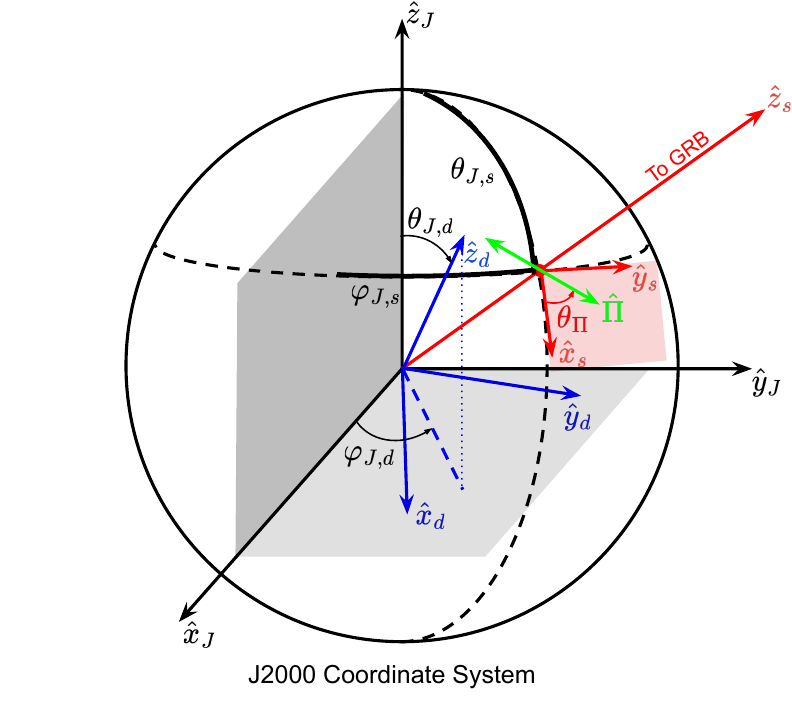} \\
\vspace{1cm}
\includegraphics[width=0.35\textwidth]{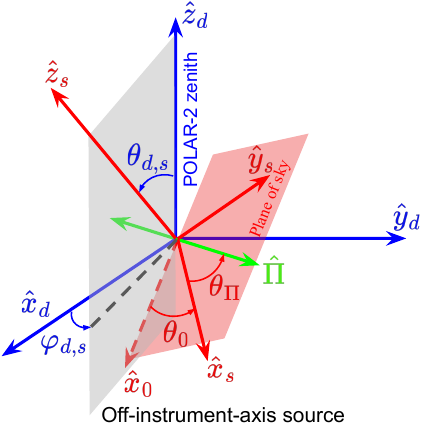}
\hspace{2cm}
\includegraphics[width=0.4\textwidth]{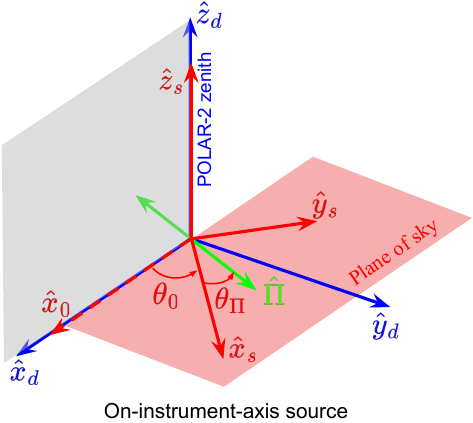}
\caption{
Coordinate Systems: (\textbf{Top}) We use a fixed J2000 coordinate system that has unit vectors $(\hat x_J,\hat y_J,\hat z_J)$. A given GRB (or source) in the direction of the radial unit vector $\hat z_s$ is localized in this system with polar angle $\theta_{J,s} = \pi/2 - \delta$, where $\delta$ is the source declination in radians, and azimuthal angle $\varphi_{J,s} =\,$\,RA, were RA is the source right ascension in radians. Two mutually orthogonal unit vectors $(\hat x_s,\,\hat y_s)$ in the direction of the polar ($\hat\theta_{J,s}$) and azimuthal ($\hat\varphi_{J,s}$) unit vectors, respectively, form the plane of the sky normal to $\hat z_s$ (red shaded region), where both vectors are tangent to the J2000 unit sphere at the location where $\hat z_s$ intersects it. The direction of $\hat x_s$ is chosen to be along the north-south great circle where $\hat x_s$ points from north to south. The plane of the sky also contains the polarization unit vector $\hat\Pi$. We report the polarization angle (PA) as $\theta_\Pi = \arccos(\hat\Pi\cdot\hat x_s)$ where it is measured in the clockwise direction from the vector $\hat x_s$ when looking along $\hat z_s$ towards the GRB. Another important coordinate system (in blue) is attached to POLAR-2, where the detector (or its zenith) points in the direction of the radial unit vector $\hat z_d$, which is localized in the J2000 system with coordinates $(\theta_{J,d},\varphi_{J,d})$. The two orthogonal sides of the square detector plane are along the unit vectors $\hat x_d$ and $\hat y_d$. 
(\textbf{Bottom-Left}) An off-instrument-axis source is localized in the detector plane with coordinates ($\theta_{d,s}$, $\varphi_{d,s}$) where the azimuthal angle is measured up to the dashed gray line which is the projection of $\hat z_s$ in the detector plane. The unit vectors $\hat z_d$ and $\hat z_s$ form a plane (gray shaded region) that intersects with the plane of the sky (red shaded region) and their intersection defines another vector $\hat x_0$ common to both planes. The azimuthal angle $\theta_0$ is measured in the plane of the sky from $\hat x_0$ up to the unit vector $\hat x_s$. 
(\textbf{Bottom-Right}) When the source is on-instrument-axis ($\theta_{d,s}=0$, to within the measurement accuracies of $\hat{z}_s$ and $\hat{z}_d$) then $\varphi_{d,s}$ is ill-defined and therefore it is fixed to $\varphi_{d,s}=0$ (i.e. $\hat{x}_0=\hat{x}_d$ in addition to $\hat{z}_s=\hat{z}_d$) and $\theta_0$ is then measured from one major axis of the detector. In this case, the detector plane coincides with the plane of the sky that contains the polarization vector.
}
\label{fig:coordinate}
\end{figure*}
%%%%%%%%%%%%%%%%%%%%%

An important tool for distinguishing between the two radiative processes and learn more about the properties of the emission region is linear polarization \citep[see][for a comprehensive discussion]{Gill+20,Gill+21}. In particular, synchrotron emission from a large-scale ordered B-field transverse to the local fluid velocity, e.g. a globally ordered toroidal field, would yield significant net polarization after integrating over the unresolved GRB image on the plane of the sky \citep{Granot-03,Lyutikov+03}. In contrast, Comptonized emission as envisaged in many photospheric emission models would yield net zero polarization in a uniform jet viewed well within its aperture, even though locally the emission is strongly polarized \citep{Beloborodov-10}. Net non-vanishing polarization in photospheric models, as well as in synchrotron models that feature axisymmetric B-fields around the local radial direction (identified with shock normal), can only be obtained in uniform jets that are viewed very close to their edges or in outflows with modest angular structure \citep{Gill+20,Ito+14,Parsotan+20}. Therefore, prompt GRB polarization is not only sensitive to the radiation mechanism, but also to the outflow angular structure, viewing geometry, and in the case of synchrotron emission the magnetic field configuration.

Measurement of $\gamma$-ray polarization has been a challenging endeavour, both on the instrumentation front as well as on the data analysis side. Several space-borne instruments, a few with properly calibrated polarimeters and most with uncalibrated ones, have taken polarization measurements  (see, e.g., Table\,1 in \citealt{Gill+21}), but no clear consensus has emerged due to large uncertainties in the results. Some success was achieved in the measurements made by POLAR, a dedicated (now defunct) GRB polarimeter mounted on the China Space Laboratory ``Tiangong-2'', that delivered relatively consistent results \citep{Kole+20}. Although it only operated for a period of six months, it performed reasonably sensitive polarization measurements of fourteen GRBs and showed that the sample containing five bursts with the most sensitive measurements tend to favour lower polarization with $\Pi\lesssim20$\,per cent. If this result is confirmed in a larger sample and with higher statistical significance, then it would strongly disfavour large scale ordered B-fields in GRB outflows (when there is clear evidence that the emission is synchrotron), and may even be used to draw important conclusions about the outflow magnetization -- another important open question. To that end, the next generation POLAR-2 instrument \citep{Kole+25} is being developed and slated for launch in the year around 2028 to be mounted on the China Space Station (CSS).

In this work, we demonstrate the improved capabilities of the POLAR-2/HPD instrument by carrying out polarimetry on synthetic dataset sourced from time and energy dependent theoretical models and convolved with the HPD's response. In doing so we develop a novel spectro-polarimetric data fitting technique that allows to constrain physical model parameters to an accuracy of a few per cent in bright GRBs. The outline of the paper is as follows. In \S\,\ref{sec:POLAR-II} we present highlights of the improved instrument design and polarimetric capabilities. We describe our time-resolved spectro-polarimetric theoretical model in \S\,\ref{sec:theory-model} and show the origin of the different model parameters. In \S\,\ref{sec:modeling} we describe our methodology of creating synthetic sources and then fitting the observations with theoretical models using MCMC to recover the initial parameters used for the synthetic source. The main findings of this work are reported in \S\,\ref{sec:results}. Finally, we summarize our results and close with some discussion in \S\,\ref{sec:discussion}.

%%%%% FIGURE %%%%%%%%%
\begin{figure*}
\centering
\includegraphics[width=0.45\textwidth]{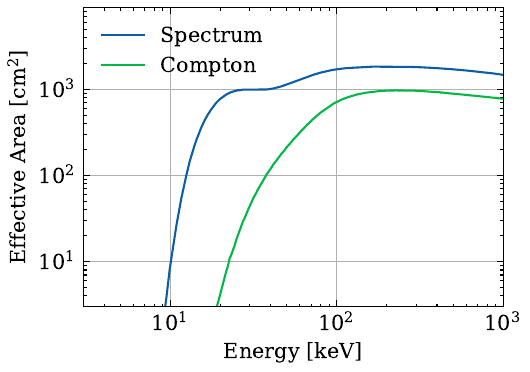}\hspace{5em}
\includegraphics[width=0.45\textwidth]{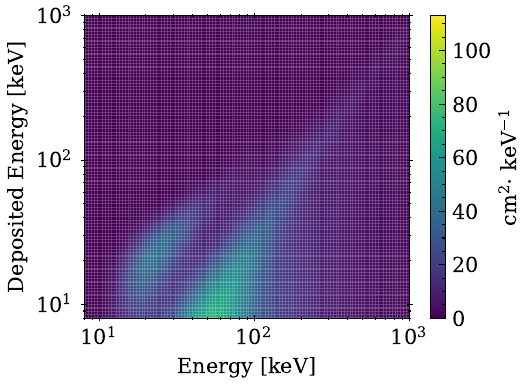}
\caption{
(\textbf{Left}) Effective area of the HPD detector of POLAR-2, shown as a function of the incident photon energy for an on-instrument-axis source with $(\theta_{d,s},\varphi_{d,s})=(0,0)$. The blue curve corresponds to the effective area for all events with energy depositions in the detectors, while the green curve shows the effective area specifically for Compton-scattering events. (\textbf{Right}) Instrument response showing the distribution of the measured (the true deposited energy cannot exceed the true energy) deposited energy as a function of the incident photon (true) energy (after \citealt{DeAngelis+23}).
}
\label{fig:eff_area}
\end{figure*}
%%%%%%%%%%%%%%%%%%%%%

%%%%%%%%%%% FIGURE %%%%%%%%%%%%%%%%%%%%%%%%%%%%%%%%%%%%%%
\begin{figure*}
\centering
\includegraphics[width=0.45\textwidth]{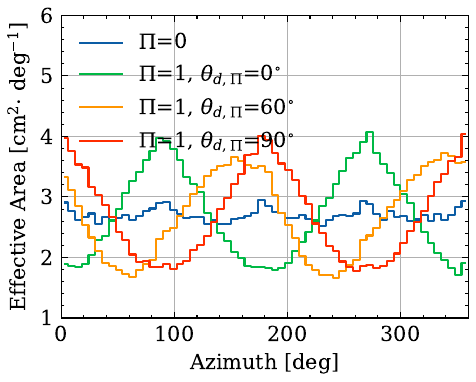}
\includegraphics[width=0.45\textwidth]{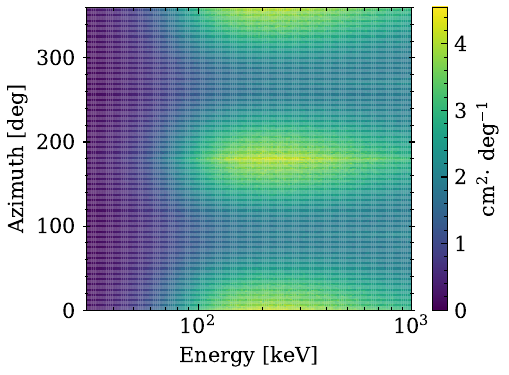}
\caption{
(\textbf{Left}) Polarimeter response showing the modulation curve or Compton scattering angle distribution for an on-instrument-axis (i.e. at the detector zenith with $\theta_{d,s}=0$) source and for $E=200$\,keV photons. Blue line shows the unpolarized case and green, yellow, and red lines show the response for a 100\% polarized source with PA $\theta_{d,\Pi} = \{0^\circ, 60^\circ, 90^\circ\}$. 
(\textbf{Right}) Polarimeter response showing the modulation curve for an on-instrument-axis source with $\Pi=1$, $\theta_{d,\Pi} = 90^{\circ}$, and photon energy $50\,\rm{keV}\leq E \leq 1000\,\rm{keV}$.
}
\label{fig:prmf}
\end{figure*}
%%%%%%%%%%%%%%%%%%%%%%%%%%%%%%%%%%%%%%%%%%%%%%%%%%%%%%%%%%

%%%%%%%%%%%%%%%%%%%%%%%%%%%%%%%%%%%%%%%%%%%%%%%%%%%%%%%
\section{POLAR-2: Instruments \& Capabilities}\label{sec:POLAR-II}
%%%%%%%%%%%%%%%%%%%%%%%%%%%%%%%%%%%%%%%%%%%%%%%%%%%%%%%
POLAR-2 \citep{Kole+25} is the successor of POLAR \citep{PRODUIT2018} and likewise an instrument dedicated to the measurement of linear polarization of prompt $\gamma$-ray photons in GRBs. It is 
slated for launch and to be mounted on the Chinese space station in the year around 2028. It will comprise three different instruments: (i) the High-energy Polarimetry Detector (HPD) will conduct $\gamma$-ray polarimetry in the $(40-1000)$\,keV energy range, (ii) the Low-energy Polarimetry Detector (LPD; \citealt{Xie+25}) will measure polarization in the $(2-10)$\, keV energy window, and (iii) the Broad-band Spectrometer Detector (BSD; \citealt{Sun+25}) will provide burst localization and spectrum in the $\sim10\,{\rm keV}-1\,{\rm MeV}$ energy range.

The method for detecting polarization of $\gamma$-ray photons in POLAR-2/HPD is Compton scattering using an array of scintillator bars. The same technique was employed in POLAR that featured 25 polarization detector units. In POLAR-2, the number of such detector units has been increased to 100 which would result in a larger effective area and greater sensitivity in detecting low levels of polarization. In measuring linear polarization this technique is realized as follows \citep[see, e.g.,][for more details]{Gill+21}. The incoming photon experiences at least one scattering in the detector, where the scattering allows to reconstruct the azimuthal angle of the scattered photon which is anti-correlated with the polarization direction of the incoming photon. The cross-section for scattering is described by the Klein-Nishina formula \citep[e.g.][]{Rybicki-Lightman-79}, which dictates that the scattering azimuth tends to be perpendicular to the direction of the polarization vector. If the source is completely unpolarized, then the scattering azimuth will no longer have this anisotropic tendency statistically and it will be axisymmetric instead. If it is partially linearly polarized, then the statistical distribution of the scattering azimuths is a linear superposition of the fully unpolarized and fully linearly polarized cases. 

The polarization of the source can be measured by comparing the azimuthal angle distribution of the Compton scattered events with the instrument response. This entails fitting the azimuthal distribution with the linear combination of the instrument response to a completely unpolarized source and completely linearly polarized source. The general method, which we describe in the next sub-section, to deal with the scattering azimuth distribution is to accumulate the observed scattering directions and divide them into bins to obtain the \textit{modulation curve}. 

%%%%%%%%%%%%%%%%%%%%%%%%%%%%%%%%%%%%%%%%%%%%%%%%%%%%%%%
\subsection{Instrument Effective Area and Response}
%%%%%%%%%%%%%%%%%%%%%%%%%%%%%%%%%%%%%%%%%%%%%%%%%%%%%%%
In order to construct the modulation curve, we first need to prepare the instrument response for a given detection. This is done using the Geant4 framework \citep{Agostinelli+03} created by the POLAR-2 instrumentation team. A given GRB is localized in the detector plane with spherical coordinates $(\theta_{d,s},\varphi_{d,s})$ as shown in the bottom two panels of Fig.\,\ref{fig:coordinate}, with subscript ``d'' indicating detector coordinates. For a given source position we run our instrument response simulation in the energy range of POLAR-2/HPD to obtain the effective area.  Figure\,\ref{fig:eff_area} shows the total effective area (blue curve) for the spectral response when the source is located at the detector zenith ($\theta_{d,s}=0$; see bottom-right panel of Fig.\,\ref{fig:coordinate}). The corresponding instrument response is shown in the right panel of Fig.\,\ref{fig:eff_area} that provides a matrix of probabilities (when normalized by the total effective area) to convert the \textit{true} energy of the incident photon into the energy measured by the detector as \textit{deposited} energy.

The polarization responses for a completely unpolarized ($\Pi=0$) and $100$\,per cent linearly polarized source ($\Pi=1$), also at the detector zenith, are generated through Compton events analysis. It provides the effective area of Compton events in the POLAR-2 energy range, as shown by the green curve in the left panel of Fig.\,\ref{fig:eff_area}, and the modulation curves, as shown in the left panel of Fig.\,\ref{fig:prmf}, for photons with energy $E=200$\,keV and different PAs on the plane of the sky. The complete modulation response over the detector's energy range is shown in the right panel of Fig.\,\ref{fig:prmf} for a fully polarized source with PA of $\theta_{d,\Pi}\equiv\theta_0+\theta_\Pi=90^\circ$.

%%%%%%%%%%%%%%%%%%%%%%%%%%%%%%%%%%%%%%%%%%%%%%%%%%%%%%%
\section{Time-Resolved Spectro-Polarimetric Model}\label{sec:theory-model}
%%%%%%%%%%%%%%%%%%%%%%%%%%%%%%%%%%%%%%%%%%%%%%%%%%%%%%%
To model the pulse profiles and temporal evolution of polarization, we consider the dynamical evolution 
of an ultrarelativistic thin-shell, having bulk Lorentz factor (LF) $\Gamma\gg1$ and lab-frame width 
$\Delta\ll R/\Gamma^2$. For high radiative efficiency during prompt emission, the radiating electrons are expected to be in the 
\textit{fast-cooling} regime, with comoving\footnote{all comoving quantities are shown with a prime} 
radiative cooling times ($t_{\rm cool}'$) much shorter than the dynamical time ($t_{\rm dyn}'=R/\Gamma c$). 
As a result, the emission arises from a very thin layer which justifies the assumption of a thin radiating 
shell. Below we follow the treatment in \citet{Gill-Granot-21} (also see \citealt{Genet-Granot-09,Uhm-Zhang-15,Uhm-Zhang-16} 
for a pulse model) and only discuss the salient points.

\subsection{Pulse Model and Spectrum}
The shell starts to radiate at radius $R=R_0$ and continues to do so until $R=R_f=R_0(1+\Delta R/R_0)$, after which point the emission shuts off. During this time, its dynamical evolution is governed by $\Gamma(R)=\Gamma_0(R/R_0)^{-m/2}$, where $\Gamma_0=\Gamma(R_0)$. The index $m$ is used to study cases in which the shell is coasting ($m=0$), accelerating ($m<0$), or decelerating ($m>0$). The spectral luminosity of the shell evolves with radius where we assume that the peak luminosity and spectral peak frequency or photon energy change with radius as a power law,
\begin{equation}\label{eq:L-and-nupk-scaling}
    L'_{\nu'}(R,\theta) = L_0'\fracb{R}{R_0}^a  S\fracb{\nu'}{\nu'_{\rm pk}}\,f(\theta)\quad{\rm with}\quad \nu'_{\rm pk} = \nu_0'\fracb{R}{R_0}^d\,,
\end{equation}
where $L_0' = L'_{\nu_{\rm pk}'}(R_0)$ and $\nu_0' = \nu_{\rm pk}'(R_0)$ are normalizations of the spectral luminosity and peak frequency at $R=R_0$. The factor $f(\theta)$ encodes the angular structure of the flow, and for a spherical flow $f(\theta)=1$. We make the simplifying assumption that the emission is isotropic in the comoving frame. In some models, anisotropic emission from electrons moving in particular directions with respect to the jet's bulk motion have also been considered \citep{Beniamini-Granot-16}. This is expected to arise in macroscopic reconnection 
events \citep{Kumar-Crumley-15}, `jets in a jet' \citep{Levinson-Eichler-93}, and relativistic turbulence scenarios \citep{Lyutikov-Blandford-03,Kumar-Narayan-09,Lazar+09}.
 
The comoving spectrum is assumed to be the phenomological Band-function \citep{Band+93}
\begin{equation}
    S(x) = e^{1+b_1} \left\{
    \begin{tabular}{c|c}
        $x^{b_1} e^{-(1+b_1)x}$\,, & $x\leq x_b$ \\
        $x^{b_2}x_b^{b_1-b_2}e^{-(b_1-b_2)}$\,, & $x\geq x_b$
    \end{tabular}\right.\,,
\end{equation}
where 
\begin{equation}
    x \equiv \frac{\nu'}{\nu_{\rm pk}'} = \frac{2\Gamma_0}{\delta_D}x_0\fracb{R}{R_0}^{-d}
    \quad{\rm with}\quad x_0\equiv\frac{\nu}{\nu_0}\,,
\end{equation}
and where $\nu_0$ is the peak frequency of the first photons emitted along the observer's LOS 
from radius $R_0$ and received at time $t=t_0$, and $\delta_D$ is the Doppler factor defined below. 
The break energy $x_b = (b_1-b_2)/(1+b_1)>1$ when $b_2<-1$. The local spectral index is given by 
\begin{equation}
    \frac{d\ln S(x)}{d\ln x} = 
    \begin{cases}
      \alpha_{\rm Band} + 1 = b_1-x(1+b_1) & x \leq x_b \\
      \beta_{\rm Band} + 1 = b_2 & x > x_b\,,
    \end{cases}
\end{equation}
where $\alpha_{\rm Band}$ and $\beta_{\rm Band}$ are the low and high energy, respectively, photon indices used in the Band function. 

How the spectral luminosity evolves with radius, in particular the value of power-law indices 
$a$ and $d$ in Eq.\,(\ref{eq:L-and-nupk-scaling}), depends on how energy is dissipated and then 
radiated in the flow. This is discussed next.

%%%%%%%%%%% FIGURE %%%%%%%%%%%%%%%%%%%%%%%%%%%%%%%%%%%%%%
\begin{figure*}
\centering
\includegraphics[width=0.95\textwidth]{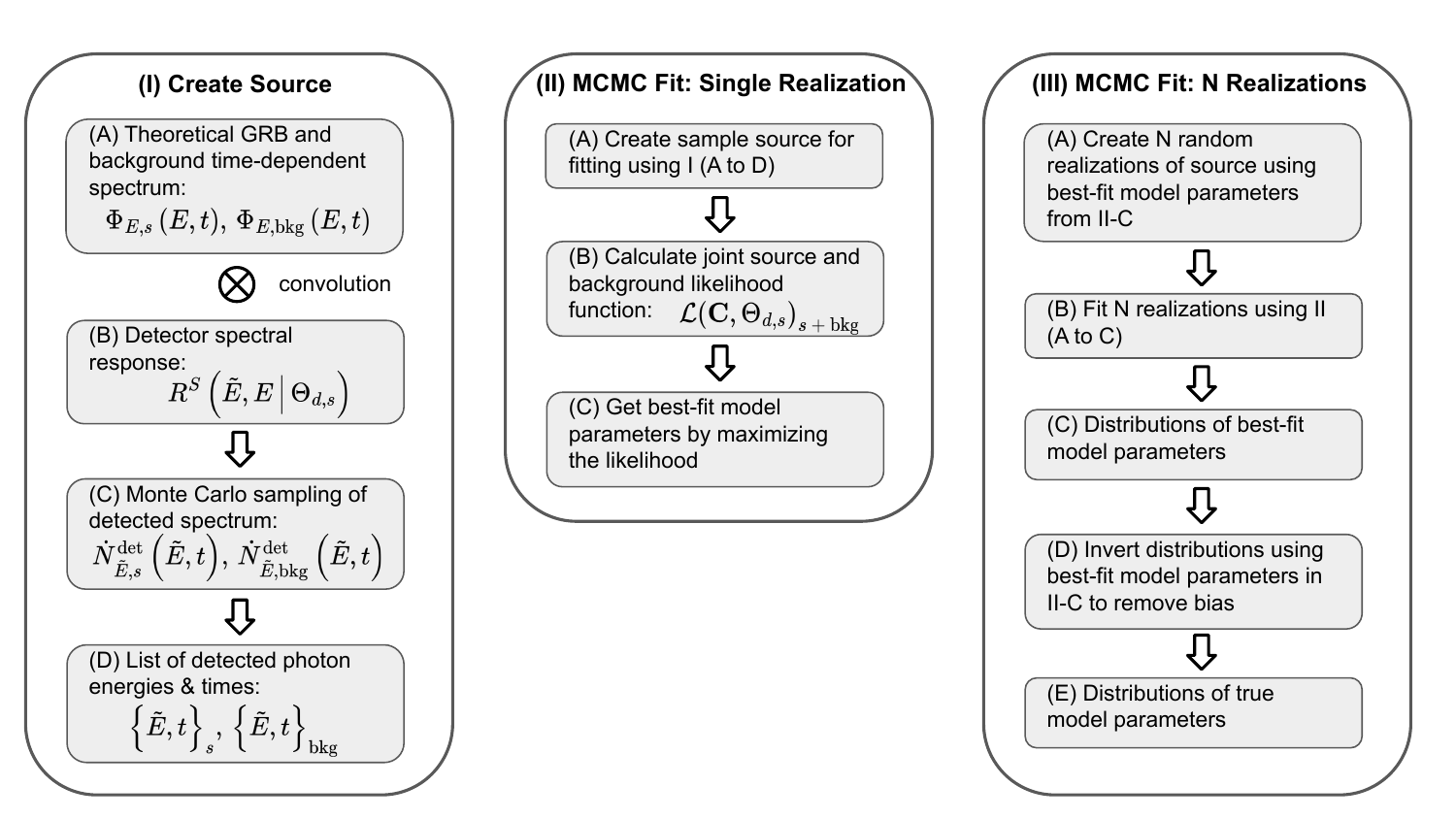}
\includegraphics[width=0.95\textwidth]{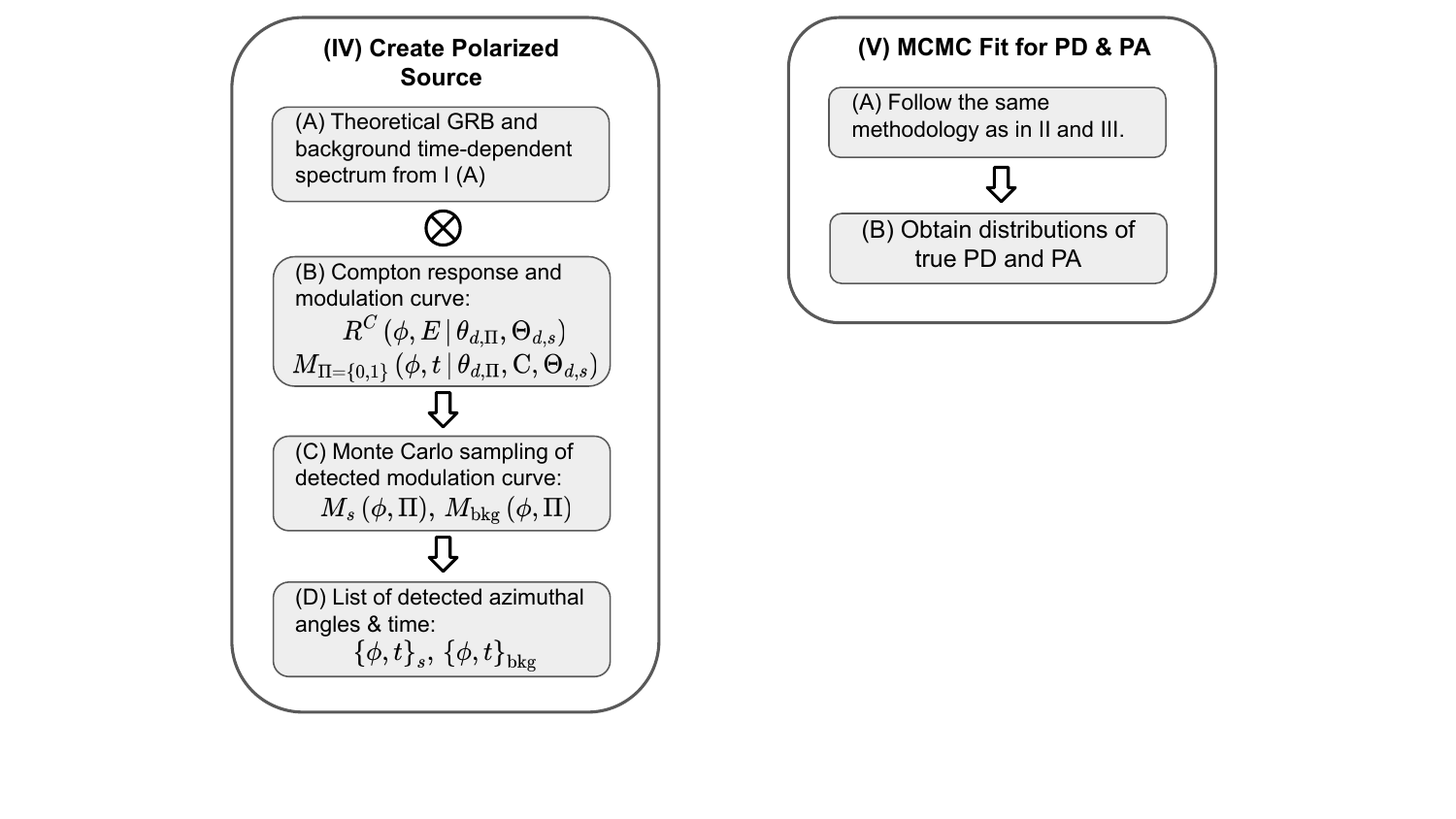}
\caption{Flow chart showcasing the adopted methodology for synthetic source creation and model fitting to the synthetic source. 
See \S\ref{sec:modeling} for details.
}
\label{fig:flowchart}
\end{figure*}
%%%%%%%%%%%%%%%%%%%%%%%%%%%%%%%%%%%%%%%%%%%%%%%%%%%%%%%%%%

%%%%%%%%%%%%%%%%%%%%%%%%%%%%%%%%%%%%%%%%%%%%%%%%%%%%%%%%%%
\subsection{Outflow Composition, Dynamics, \& Radiation}
%%%%%%%%%%%%%%%%%%%%%%%%%%%%%%%%%%%%%%%%%%%%%%%%%%%%%%%%%%

The dynamical evolution of the outflow, and how energy is dissipated and then ultimately radiated, are sensitive to its composition, typically expressed in terms of the flow magnetization. For a 
proton-electron plasma (neglecting any $e^\pm$-pair enrichment), the magnetization $\sigma = w'_B/w'_m = B'^2/4\pi n'm_pc^2$ is given by the ratio of the magnetic field enthalpy density, $w'_B = B'^2/4\pi$, to that of matter (assuming cold protons and electrons), $w'_m = n'm_pc^2$. Here $n'$ is the particle number density, $B'$ is the magnetic field strength, $m_p$ is the proton mass, and $c$ is the speed of light. When $\sigma\ll1$, most of the energy resides in the kinetic energy of the baryons, therefore making the flow kinetic-energy-dominated (KED). Part of this energy is dissipated in internal shocks and then part of that is radiated. Alternatively, when $\sigma>1$, the outflow is Poynting flux dominated (PFD) and the main energy reservoir is the magnetic field, in which case energy is dissipated via magnetic reconnection and/or magneto-hydrodynamic (MHD) instabilities. The dynamics of the flow are different in both scenarios.

In the canonical internal shocks model \citep{Rees-Meszaros-94,Paczynski-Xu-94,Sari-Piran-97,Daigne-Mochkovitch-98}, the central engine accretes intermittently and ejects shells of matter that are initially separated by length scale $\sim ct_v/(1+z)$ and have fluctuations in bulk LFs of order $\Delta\Gamma\sim\Gamma$, i.e. the mean bulk LF of the unsteady flow. As a result, after the outflow acceleration saturates with $\Gamma(R)=\Gamma_\infty\propto R^0$, and therefore $m=0$, faster moving shells catch up from behind with slower ones and collide with each other to dissipate their kinetic energy at internal shocks. 

Alternatively, a PFD outflow is permeated by strong magnetic fields advected from the base of the flow \citep[e.g.,][]{Thompson-94,Lyutikov-Blandford-03}. This renders internal shocks inefficient at dissipating kinetic energy. Instead, magnetic reconnection and/or MHD instabilities, e.g. the Kruskal-Schwarzchild instability \citep{Lyubarsky-10b,Gill+18} which is the magnetic analog of the Rayleigh-Taylor instability, become viable energy dissipation mechanisms. A popular model of a PFD outflow is that of a striped-wind \citep{Lyubarsky-Kirk-01,Spruit+01,Drenkhahn-02,Drenkhahn-Spruit-02,Begue+17} where the magnetic field lines reverse polarity within the flow, and magnetic energy is dissipated when opposite polarity field lines are brought together. A significant fraction of the dissipated energy goes towards accelerating the flow with $\Gamma\propto R^{1/3}$, which therefore yields $m=-2/3$. Broadly similar flow dynamics is obtained in a highly variable magnetized outflow \citep{Granot+11,Granot-12,Komissarov12} that doesn't require magnetic field polarity reversals to dissipate energy in mildly magnetized internal shocks with the peak dissipation radius near the coasting radius, similar to a striped wind.

To understand the radial dependence of the spectral luminosity and peak frequency, a prescription of the radiation mechanism is needed. Here we focus on synchrotron emission from shock accelerated relativistic power-law electrons. The alternative is Comptonized emission from mildly relativistic electrons. Both radiation mechanisms have been shown to produce Band-like prompt GRB spectrum \citep[e.g.,][]{Gill+20b}. A complete description of the radial dependence of synchrotron emission follows from the scaling laws of the lab-frame magnetic field, $B(R)\propto R^\ell$, and that of the minimal LF of the power-law electrons, $\gamma_m(R)\propto R^s$. By using these scalings it can be shown (see \citealt{Gill-Granot-21} for more details) that the ($\nu F_\nu$) peak frequency scales with radius as $\nu_{\rm pk}'(R)\propto R^{(\ell+2s+m/2)}$ and the spectral luminosity at the peak frequency scales as $L'_{\nu'_{\rm pk}}(R)\propto R^{-(\ell+s)}$.

In a KED flow, collisions between different mass shells give rise to a two shock structure comprising forward and reverse shocks that dissipate the initial kinetic energy of the shells \citep[e.g.][]{Rahman+24}. The strength of the two shocks depends on the relative bulk LF of the upstream and downstream material ($\Gamma_{ud}$), which remains constant with radius for the simplest case of uniform shells. As a result, $\gamma_m\propto\Gamma_{ud}-1\propto R^0$ and therefore $s=0$. Internal shocks take place at a radius much larger than that where the initial acceleration of the flow has saturated and the flow starts to coast ($m=0$). Magnetic flux conservation in a coasting shell with fixed lab-frame width then leads to dilution of the radial component of the magnetic field, 
with $B_r(R)\propto R^{-2}$, at a rate much faster than its transverse component, $B_{\theta,\varphi}(R)\propto R^{-1}$. Consequently, at large distances the transverse components dominates, which means that $\ell=-1$. Putting it all together, we find that the power-law indices in Eq.\,\ref{eq:L-and-nupk-scaling} are $a_{\rm KED}=1$ and $d_{\rm KED}=-1$.

In a PFD flow, the magnetic field can be axisymmetric, e.g. a globally toroidal field centered on the jet symmetry axis. 
In that case, the poloidal component declines faster with radius, with $B_p\propto R^{-2}$, as compared to the toroidal 
component that scales as $B_\phi\propto R^{-1}$. Therefore, at large distances from the central engine, where magnetic 
dissipation occurs, the toroidal component dominates, which again yields $\ell=-1$. The scaling for $\gamma_m$ is obtained 
from the mean energy per unit rest mass of the power-law electrons, which in turn depends 
on the magnetization and its radial dependence, such that $\gamma_m\propto\langle\gamma_e\rangle\propto\sigma\propto R^{2+2\ell+m/2}$, 
which yields $s=-1/3$ \citep{Gill-Granot-21}. This finally gives $a_{\rm PFD}=4/3$ and $d_{\rm PFD}=-2$.

%%%%%%%%%%%%%%%%%%%%%%%%%%%%%%%%%%%%%%%%%%%%%%%%%%%%%%%
\subsection{Magnetic Field \& Jet Structure}\label{sec:B-field-jet-structure}
%%%%%%%%%%%%%%%%%%%%%%%%%%%%%%%%%%%%%%%%%%%%%%%%%%%%%%%

To calculate the linear polarization, we need to know the structure of the magnetic field in the emission region. Here 
we consider four physically motivated structures \citep[e.g.,][]{Gill+20}:
\begin{enumerate}
    \item $B_{\rm ord}$: An ordered B-field in the plane of the ejecta having a coherence length angular scale as large as 
    that of the beaming cone, such that $\theta_B\gtrsim\Gamma^{-1}$ \citep{Granot-03}.
    \item $B_\perp$: A shock-generated tangled (randomly oriented) B-field with $\theta_B\ll\Gamma^{-1}$ constrained to be 
    in the plane transverse to the local velocity vector, which we assume to be in the radial direction \citep{Granot-03}.
    \item $B_\parallel$: An alternative to the previous case and generalization of the shock-generated field, where the field 
    is now ordered but aligned with the local velocity vector \citep{Granot-03}.
    \item $B_{\rm tor}$: An axisymmetric globally ordered toroidal field expected to arise in PFD outflows \citep{Lyutikov+03,Granot-Taylor-05}.
\end{enumerate}

The net polarization is not only sensitive to the B-field structure but also to the viewing angle as well as the jet angular structure. For example, the net polarization vanishes for a spherical flow when the B-field is axisymmetric around the LOS, e.g. $B_\perp$ or $B_\parallel$, due to complete cancellation of polarization vectors that are symmetric around the LOS of the observer. This symmetry is naturally broken when the B-field is anisotropic around the LOS, e.g. in $B_{\rm ord}$ and $B_{\rm tor}$ (except when the viewing angle is $\theta_{\rm obs}=0$). Alternatively, the symmetry is broken when the jet has angular structure. The extreme case of which is when the LOS passes close to the edge of a uniform jet, as in a top-hat jet, with $\theta_{\rm obs}\gtrsim\theta_j+\Gamma^{-1}\Longleftrightarrow q\equiv\theta_{\rm obs}/\theta_j\gtrsim 1+\xi_j^{-1/2}$ where $\theta_j$ is the angular size of the jet aperture and $\xi_j = (\Gamma\theta_j)^2$.

%%%%%%%%%%%%%%%%%%%%%%%%%%%%%%%%%%%%%%%%%%%%%%%%%%%%%%%
\subsection{Model Parameters} \label{sec:model-params}
%%%%%%%%%%%%%%%%%%%%%%%%%%%%%%%%%%%%%%%%%%%%%%%%%%%%%%%
When combining the outflow dynamics, spectrum, B-field, and jet structure profiles, it results in a total of 13 model parameters: $m$, $a$, $d$, $\Delta R/R_0$, $b_1$, $b_2$, $\nu_0$, $t_0$, $q$, $\xi_j$, the source redshift $z$, B-field type, and normalization of the spectrum given by source fluence $\mathcal{F}$. For sufficiently bright events the redshift is usually determined. The dynamical and spectral parameters can be most robustly determined by forward-folding a given model (either phenomenological or physical) using the detector response and then comparing the model spectrum at a given time with the observed one in count space, as we demonstrate below. The theoretical pulse profiles and time-resolved polarization curves that are used in this work, exploring different values of the given parameters, are presented in \citet{Gill-Granot-21}.

%%%%%%%%%%%%%%%%%%%%%%%%%%%%%%%%%%%%%%%%%%%%%%%%%%%%%%%
\section{Model Fit to Synthetic Sources}\label{sec:modeling}
%%%%%%%%%%%%%%%%%%%%%%%%%%%%%%%%%%%%%%%%%%%%%%%%%%%%%%%
Here we describe the creation of a ``test case'' that is a synthetic source prepared using the time-dependent spectro-polarimetric theoretical model as presented above. The test case comprises complete spectral and polarization evolution of the emission over a single pulse or, in a more complex scenario, over multiple overlapping pulses. The source position and relative orientation of the PA in the detector plane are additional degrees of freedom introduced in the synthetic source creation. The forward-folded (after convolving with the detector response) test case lightcurve, spectra, and polarization are then fitted to the forward-folded theoretical models to constrain a selected set of model parameters. The flowchart in Fig.\,\ref{fig:flowchart} describes the entire process and it is complemented by Fig.\,\ref{Fig:lc-spec-fit-demo} that graphically shows the different products obtained in this exercise. We refer to these figures in what follows.

%%%%%%%%%%%%%%%%%%%%%%%%%%%%%%%%%%%%%%%%%%%%%%%%%%%%%%%%%%
\subsection{Synthetic Source Creation: The Test Case}
%%%%%%%%%%%%%%%%%%%%%%%%%%%%%%%%%%%%%%%%%%%%%%%%%%%%%%%%%%

\begin{landscape}
\begin{figure}
    \centering
    \includegraphics[width=0.4\textwidth]{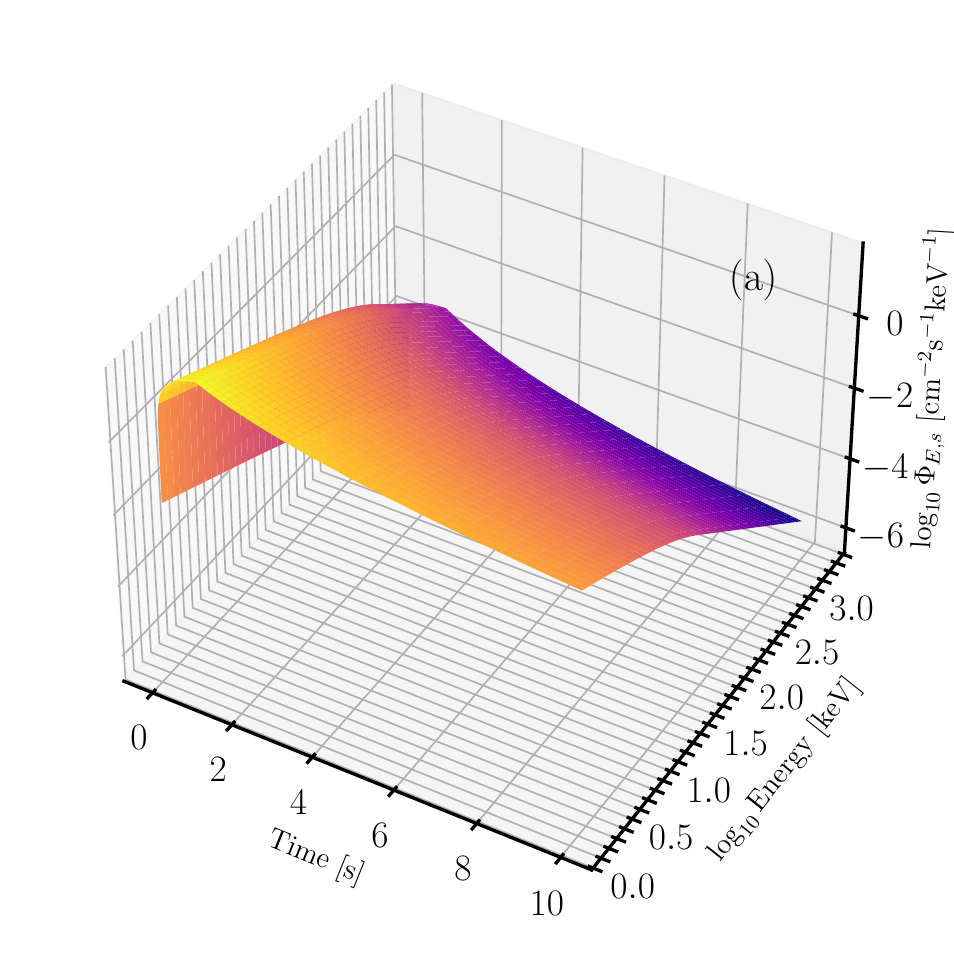}
    \quad\quad
    \includegraphics[width=0.4\textwidth]{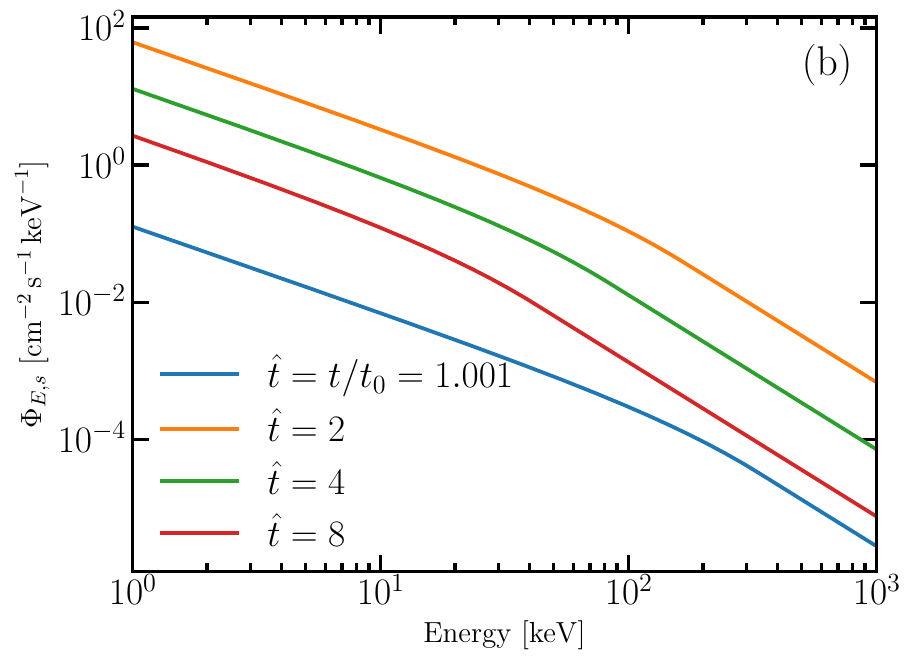}
    \quad\quad
    \includegraphics[width=0.4\textwidth]{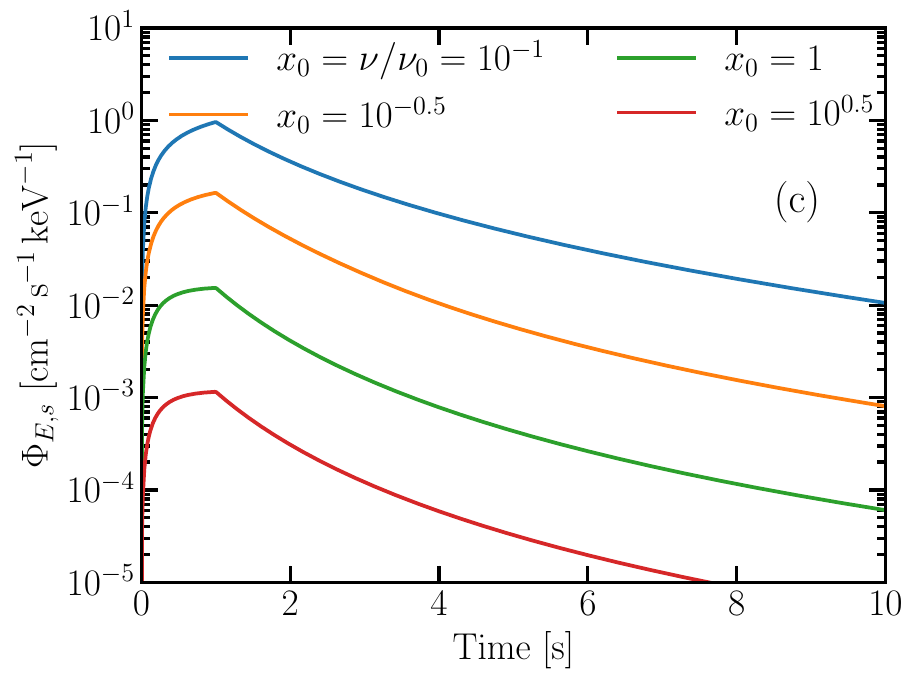}
    \quad\quad
    \\
    \includegraphics[width=0.4\textwidth]{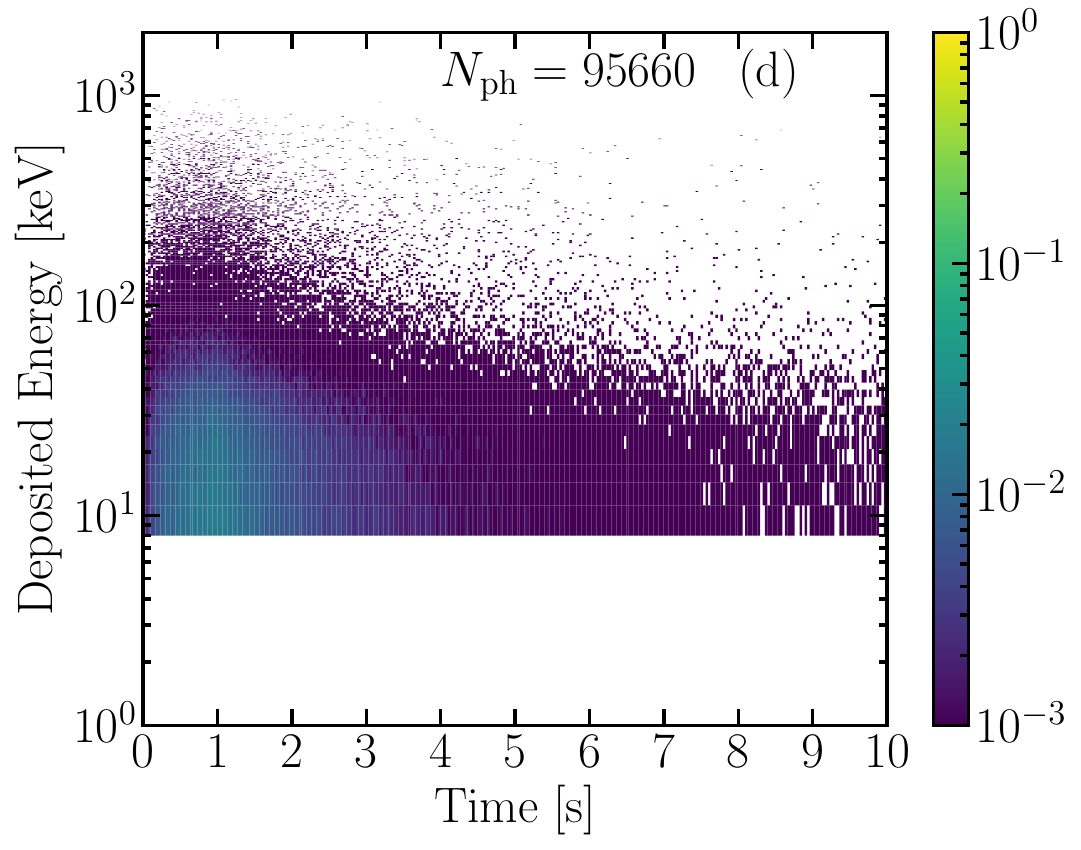}
    \quad\quad
    \includegraphics[width=0.4\textwidth]{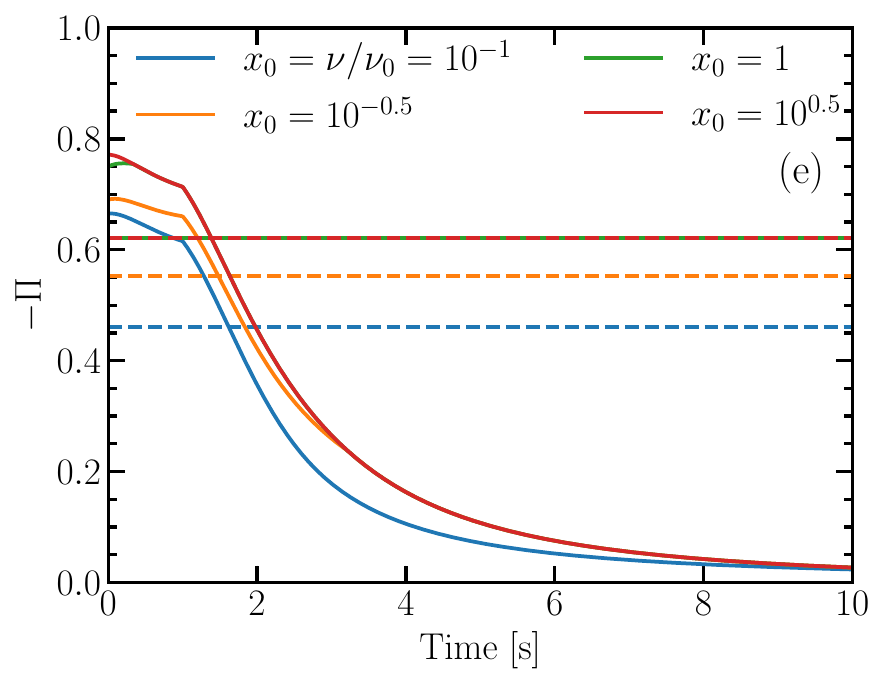}
    \quad\quad
    \includegraphics[width=0.4\textwidth]{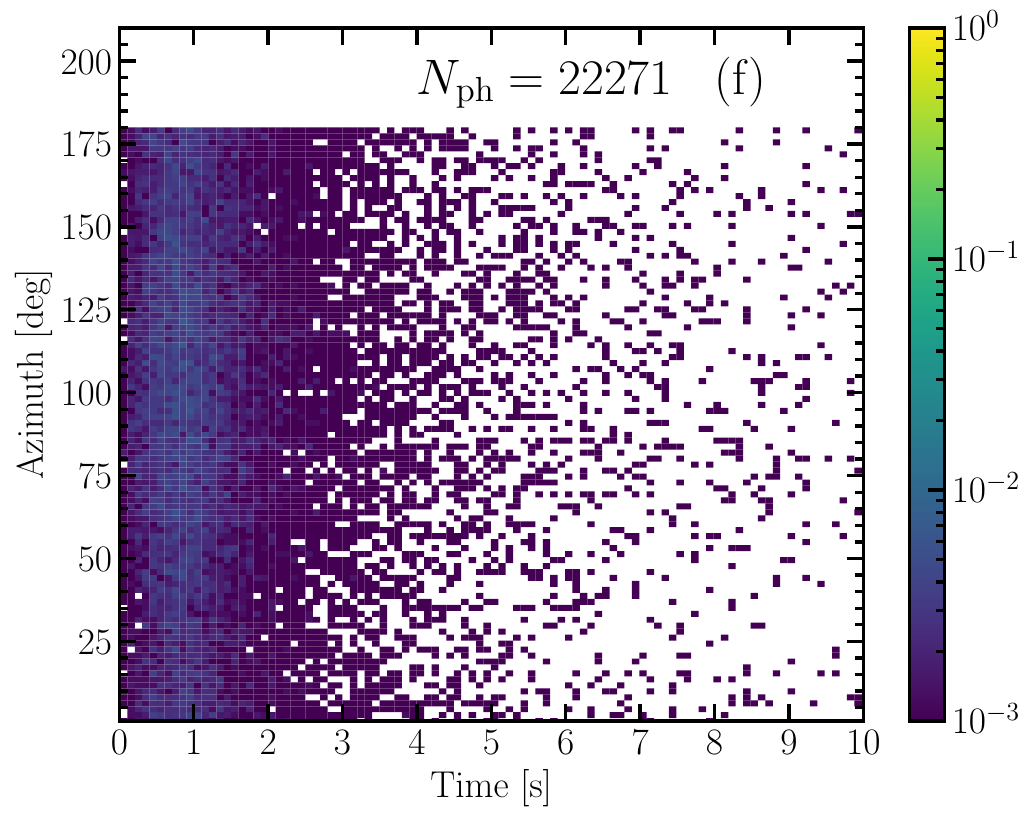}
    \caption{ 
    (\textbf{a}) Incident photon spectrum $\Phi_{E,s}(E,t)$ two-dimensional surface in the energy and time plane for the following set of model parameters: 
    $t_0=1\,$s, $\nu_0=250\,$keV, $b_1=-0.25$, $b_2=-1.25$, $a=1$, $d=-1$, $m=0$, $\Delta R/R_0=1$, $\xi_j=10^2$, $q=0$, $\mathcal{F}=10^{-5}\,{\rm erg\,cm}^{-2}$.
    (\textbf{b}) Photon spectrum at different normalized times, $\hat t = t/t_0$. 
    (\textbf{c}) Lightcurves at different normalized energies, $x_0 = \nu/\nu_0$. 
    (\textbf{d}) Distribution of detected photons in deposited energy and arrival time, obtained by sampling a single realization of the detected photon spectrum. $N_{\rm ph}$ shows the number of sample photons. 
    \confirm{(\textbf{e}) Polarization curves from an ordered B-field at different $x_0$ values. Dashed lines show the pulse-integrated polarization.} 
    (\textbf{f}) Distribution of Compton detected photons in scattering azimuth and arrival time, obtained by sampling a single realization of the detected modulation curve. 
    }
    \label{Fig:lc-spec-fit-demo}
\end{figure}
\end{landscape}

To create a test case we consider a GRB in the detector plane (see Fig.\,\ref{fig:coordinate}) with spherical coordinates $(\theta_{d,s}, \varphi_{d,s})$ and photon number spectrum $\Phi_{E,s}\equiv dN_s/dAdtdE$ (point I.\,A in Fig.\,\ref{fig:flowchart}), which gives the number of \textit{source} photons $dN_s$ detected per unit observer time $dt$, over an infinitesimal area $dA$, and per unit energy $dE$, as shown in panel (a) of Fig.\,\ref{Fig:lc-spec-fit-demo}. The photon number spectrum, $\Phi_{E,s} = F_E/E$, is obtained from the un-normalized flux density, $F_E$, that is provided by the theoretical model for a given set of model parameters. To normalize the photon spectrum we fix the fluence, which is given by $\mathcal{F} = \int dt \int dE F_E$ over the energy range of the instrument with $E_{\min}\leq E \leq E_{\max}$ and the duration of the GRB with $0\leq t\leq t_{\rm GRB}$. Here we particularly choose the energy range of \textit{Fermi}-GBM, with $10\,{\rm keV}\leq E \leq 1\,{\rm MeV}$, to normalize the photon spectrum for a given fluence to aid comparison with GRBs detected by \textit{Fermi}-GBM.

A time-dependent model spectrum for a fluence of $\mathcal{F}=10^{-5}\,{\rm erg\,cm}^{-2}$ is shown in the top three panels (a--c) of Fig.\,\ref{Fig:lc-spec-fit-demo}. Panel (a) shows the spectral surface over energy and time, and panels (b) and (c) show the spectra at different times and lightcurves at different normalized energies, respectively.

Next, we prepare the detector's spectral response (I.\,B), $R^S(\tilde E, E\vert \Theta_{d,s})=dA_{\rm eff}(\tilde E,E\vert\Theta_{d,s})/d\tilde E\equiv R^S(\tilde E,E)$ for a given source location in the detector plane with $\Theta_{d,s}\equiv(\theta_{d,s},\varphi_{d,s})$, as shown in Fig.\,\ref{fig:eff_area}. We then obtain the detected photon spectrum (I.\,C) by taking a convolution of the incident source spectrum with this response to obtain
\begin{equation}
    \dot N_{\tilde E,s}^{\rm det}(\tilde E,t) = \frac{dN_s}{dt d\tilde E} 
    = \int dE\,\Phi_{E,s}(E,t) R^S(\tilde E,E)\,.
\end{equation}
The instrument effective area, $A_{\rm eff}(E\vert\Theta_{d,s})=\int d\tilde E\,R^S(\tilde E,E)$, depends on the energy $E$ of the incident photon and the source off-instrument-axis angle $\theta_{d,s} = \arccos(\hat z_d\cdot\hat z_s)$, where $\hat z_d$ is the direction of the unit vector along which the detector is pointing and $\hat z_s$ is the unit vector pointing towards the source GRB (see Fig.\,\ref{fig:coordinate}). The detector's spectral response, $R^S(\tilde E, E)$, when normalized by the effective area at energy $E$ of the incident photon yields the probability that the incident photon will deposit energy $\tilde E$ for a given $E$. The total number of detected \textit{source} photons, $N_s^{\rm det}=\kappa N_s$ where $\kappa<1$, is then obtained from 
\begin{equation}\label{eq:Nph-det}
    N_s^{\rm det} = \int_{0}^{t_{\rm GRB}} dt \int_{\tilde E_{\min}}^{\tilde E_{\max}} 
    d\tilde E\,\, \dot N_{\tilde E,s}^{\rm det}(\tilde E,t)\,,
\end{equation}
where the instrument detects photons over the energy range $\tilde E_{\min}\leq \tilde E \leq \tilde E_{\max}$. More accurately, this is the expected or mean value of $N_s^{\rm det}$, while in practice $N_s^{\rm det}$ will be drawn from a Poisson distribution with this mean value, but here for simplicity we will neglect this distinction. 

We can now proceed to sample the detected time-dependent photon spectrum using Monte Carlo rejection sampling. Here we briefly summarize the steps for this procedure. First, we randomly draw the detected photon energy $\tilde E \in \{\tilde E_{\min}, \tilde E_{\max}\}$ and the photon arrival time $t \in \{0, t_{\rm GRB}\}$ from a uniform distribution. To sample the photon spectrum we draw another random number $0\leq \mathcal N\leq \dot N_{\tilde E,s,\max}^{\rm det}$ distributed uniformly, where $N_{\tilde E,s,\max}^{\rm det}$ is the maximum of the two dimensional surface shown in panel (a) of Fig.\,\ref{Fig:lc-spec-fit-demo}. When $\mathcal{N}\leq \dot N_{\tilde E,s}^{\rm det}(\tilde E,t)$ the photon is added to the list of detected photons and otherwise it is discarded. This procedure is repeated until the number of detected photons reaches $N_s^{\rm det}$ that yields a list of source photons with arrival times and deposited energies, $\{\tilde E_i, t_i\}_s$ (I.\,D). The distribution of the sampled photons in both energy and time is shown in panel (d) of Fig.\,\ref{Fig:lc-spec-fit-demo}. This shows a sharp cutoff at $\tilde E = 8$\,keV below which the effective area of the detector drops sharply.

%%%%%%%%%% FIGURE %%%%%%%%%%%%%%%%%%%%%%%%%%%%%%%%%%%%%%%%%%%%%%%%%%%%%%%%
%%%%%%%%%% FIGURE %%%%%%%%%%%%%%%%%%%%%%%%%%%%%%%%%%%%%%%%%%%%%%%%%%%%%%%%

%%%%%%%%%% FIGURE %%%%%%%%%%%%%%%%%%%%%%%%%%%%%%%%%%%%%%%%%%%%%%%%%%%%%%%%
\begin{figure*}
\centering
\includegraphics[width=0.45\textwidth]{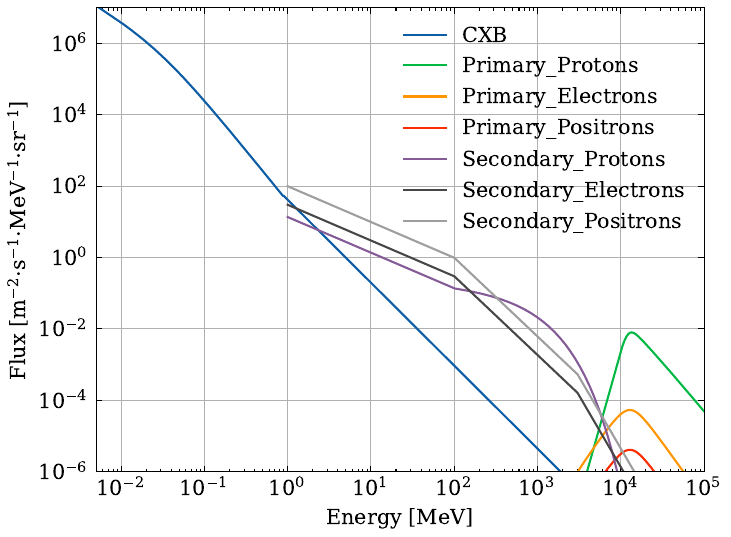}
\quad\quad
\includegraphics[width=0.45\textwidth]{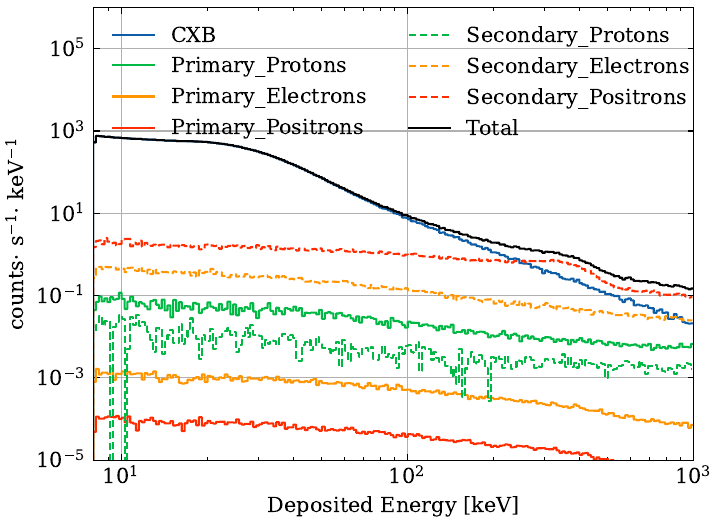}
\caption{
    (\textbf{Left}) The input energy spectrum used in the background simulation, assuming each background component is isotropic and time-invariant. See \S\ref{sec:modeling} for details. 
    (\textbf{Right}) The simulated detected background energy spectrum of POLAR-2/HPD, showing the dominant contribution from the Cosmic X-ray background (CXB).
    }
\label{fig:backgorund_input}
\end{figure*}
%%%%%%%%%% FIGURE %%%%%%%%%%%%%%%%%%%%%%%%%%%%%%%%%%%%%%%%%%%%%%%%%%%%%%%%

%%%%%%%%%%%%%%%%%%%%%%%%%%%%%%%%%%%%%
\subsection{Source Background}
%%%%%%%%%%%%%%%%%%%%%%%%%%%%%%%%%%%%%
Since POLAR-2 will be placed in a low Earth orbit on board the CSS, there are different sources that contribute to the detector background. The left panel of Fig.\,\ref{fig:backgorund_input} presents the primary environmental background components for low-Earth orbit satellites. The cosmic X-ray background (CXB) makes the dominant contribution below an energy of 1\,MeV, where the CXB can be described by a smoothly broken power-law spectrum \citep{GEHRELS+92}. Above that energy, the cosmic-ray particles dominate the background and their energy spectra are adopted from the measurements of the Alpha Magnetic Spectrometer \citep[AMS;][]{Alcaraz+00,Alcaraz+01,Mizuno+07}. By using the energy spectrum from Fig.~\ref{fig:backgorund_input} as input, the HPD detector's count spectrum can be obtained through Geant4 simulation, as shown in the right panel of Fig.\,\ref{fig:backgorund_input}. The simulated background data spans over 1000 seconds to obtain sufficient statistics. The detected background spectra are smoothed and normalized by time, resulting in the background detection model $\dot N_{\tilde E,\rm bkg}^{\rm det}(\tilde E) = dN_{\rm bkg}^{\rm det}/d\tilde E dt$. \confirm{The Earth albedo background is not included in our model, in contrast to the case in \citep{biltzinger2020,ajello2008}. This is because POLAR-2/HPD always points toward the zenith, keeping the Earth permanently outside its field of view. Therefore, any direct contribution from Earth albedo is negligible. However, high-energy protons in the South Atlantic Anomaly (SAA) can activate the detector and surrounding platform materials, producing a delayed background that decays exponentially after the detector exits the SAA and may also include characteristic activation lines. These effects will be accounted for in detail in future analysis.}

In this work, we simply assume a constant background count rate, owing to the complexity of accounting for its variations with charged particles, the SAA, solar flares, and other factors. The total number of detected background photons over duration $\Delta t = t_{\rm GRB}$ is
\begin{eqnarray}\label{eq:Nph-bkg}
    N_{\rm bkg}^{\rm det} &=& \int_{0}^{t_{\rm GRB}} dt \int_{\tilde E_{\min}}^{\tilde E_{\max}} 
    % d\tilde E\,\, \dot N_{\tilde E}^{\rm bkg}(\tilde E)\,\cdot(t_{\rm max}-t_{\rm min}).
    d\tilde E\,\, \dot N_{\tilde E,\rm bkg}^{\rm det}(\tilde{E},t)
    \\ \nonumber
    &\to&\;
    \Delta t \int_{\tilde E_{\min}}^{\tilde E_{\max}} d\tilde E\,\, \dot N_{\tilde E,\rm bkg}^{\rm det}(\tilde{E})\,.
\end{eqnarray}
Monte Carlo sampling is applied to the detected background spectrum to obtain a list of background photons, which are then added to the list of source photons. 

This concludes the preparation of the test source. We first fit the dynamical and spectral model parameters in the next section before fitting the polarized emission.

%%%%%%%%%%%%%%%%%%%%%%%%%%%%%%%%%%%%%
\subsection{Model Fitting to Unbinned Spectral Data}
\label{sec:lc-spectral-fit-method}
%%%%%%%%%%%%%%%%%%%%%%%%%%%%%%%%%%%%%
Here we develop a procedure for maximum likelihood estimation while performing an unbinned fit, first over the spectrum and lightcurve and then over the polarized emission. Our general method that is also applied later for the polarized emission is as follows. Let $\textbf{C}=\left\{C_j\right\}_{j=1}^{m}$ be the model parameters whose values we wish to constrain through the fit to the data. The data consists of $N_s^{\rm det} + N_{\rm bkg}^{\rm det}$ detected photons with measured arrival times and detected energies. The model photon spectrum of detected photons is given by $\dot N_{\tilde E,s}^{\rm det}(\tilde E,t|\mathbf C,\Theta_{d,s}) + \dot N_{\tilde E,\rm bkg}^{\rm det}(\tilde E)$ for a given set of model parameters $\textbf{C}$ and source location $\Theta_{d,s}$ in the detector plane. This is the un-normalized probability density function (PDF; per unit energy and time) for a photon of given detected energy $\tilde E$ arriving at a given time $t$. To normalize the PDF we calculate the total number of detected photons using Eq.\,\ref{eq:Nph-det} \& \ref{eq:Nph-bkg}, such that the probability for a single $i^{\rm th}$ photon of detected energy $\tilde E_i$ at time $t_i$ is
\begin{equation}
    P_i(\tilde E_i,t_i|\mathbf C,\Theta_{d,s}) 
    = \frac{\dot N_{\tilde E,s}^{\rm det}(\tilde E_i,t_i|\mathbf C,\Theta_{d,s})+
    \dot N_{\tilde E,\rm bkg}^{\rm det}(\tilde E)}{N_s^{\rm det}+N_{\rm bkg}^{\rm det}}\ .
\end{equation}
Using this probability we can define the log-likelihood, 
\begin{equation}\label{eq:log-likelihood}
    \log\mathcal{L}(\textbf{C},\Theta_{d,s}) = \sum_{i=1}^{N_s^{\rm det}+N_{\rm bkg}^{\rm det}}\log P_i(\tilde E_i,t_i|\textbf{C},\Theta_{d,s})\,, 
\end{equation}
which we maximize using Markov Chain Monte Carlo to obtain the posterior distributions of our model parameters \textbf{C} (points II and III in Fig.\,\ref{fig:flowchart}; see section\,\ref{sec:results} for further details). 

The likelihood also depends on the source position in the detector plane. This can be determined to some accuracy by using the positional information provided by other instruments, e.g. the IPN, which will provide the source coordinates (RA, Dec) with positional errors on the celestial sphere. Then, given the direction towards which the instrument is pointing, the source coordinates ($\theta_{d,s},\varphi_{d,s}$) in the detector plane along with the uncertainties can be obtained. Additionally, the dedicated spectrometer BSD of POLAR-2 onboard CSS provides sub-degree localization precision through coded-aperture mask imaging technology (\citealt{Sun+25}). This localization capability will significantly increase the sample size of GRB polarimetry measurements by POLAR-2, enabling synergistic observations with HPD and LPD without reliance on external GRB localization instruments.

The above procedure will yield the best-fit dynamical and spectral model parameters that describe the time-dependent spectral evolution of the prompt emission over the duration of a single pulse. The same can be used to fit the model over more complicated pulse shapes, including several overlapping pulses. 

%%%%%%% FIGURE %%%%%%%%%%%%%%%%%%%%%%%%%%%%%%%%%%%%%%%%%%%%%
\begin{figure*}
    \centering
    \includegraphics[width=0.7\textwidth]{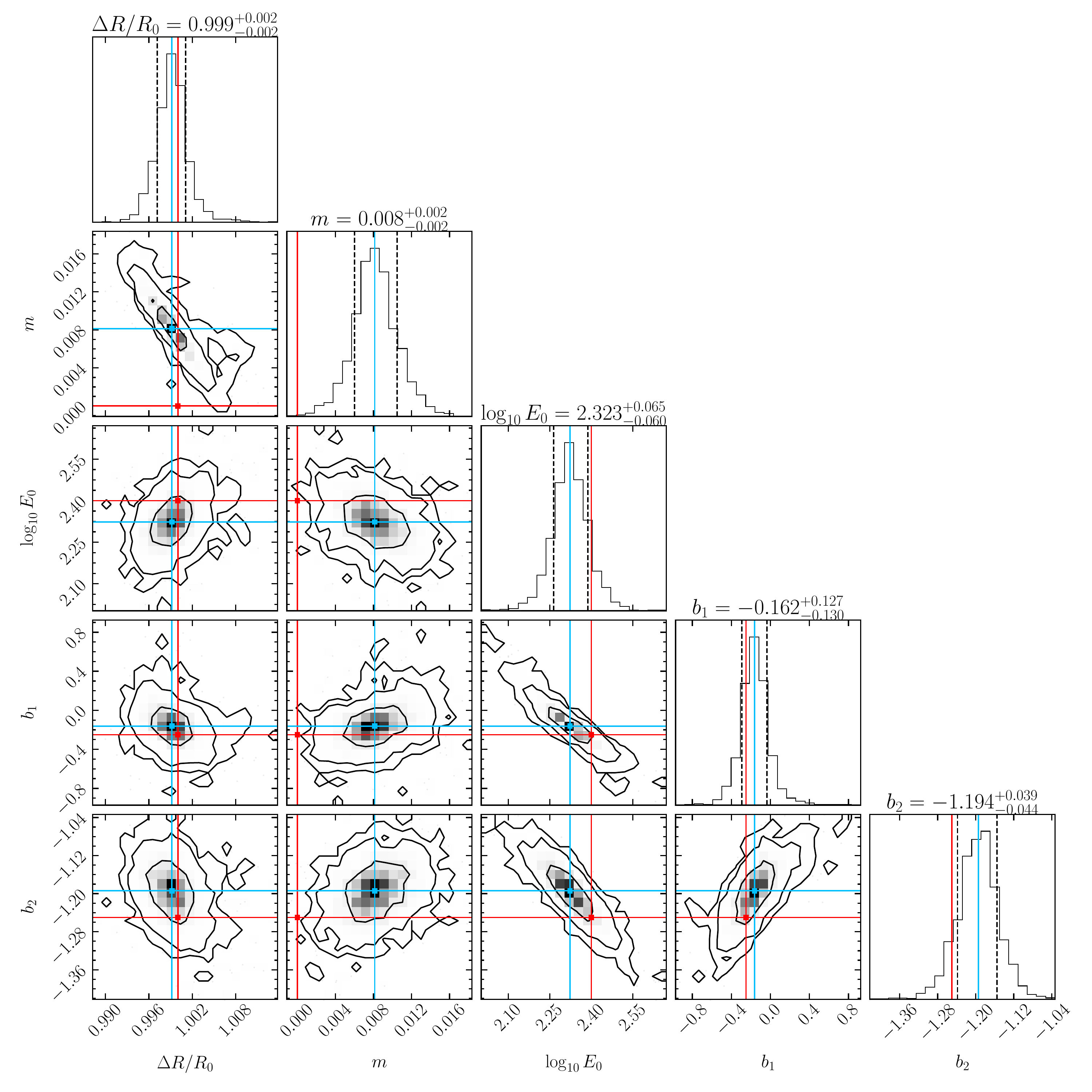}
    \includegraphics[width=0.45\textwidth]{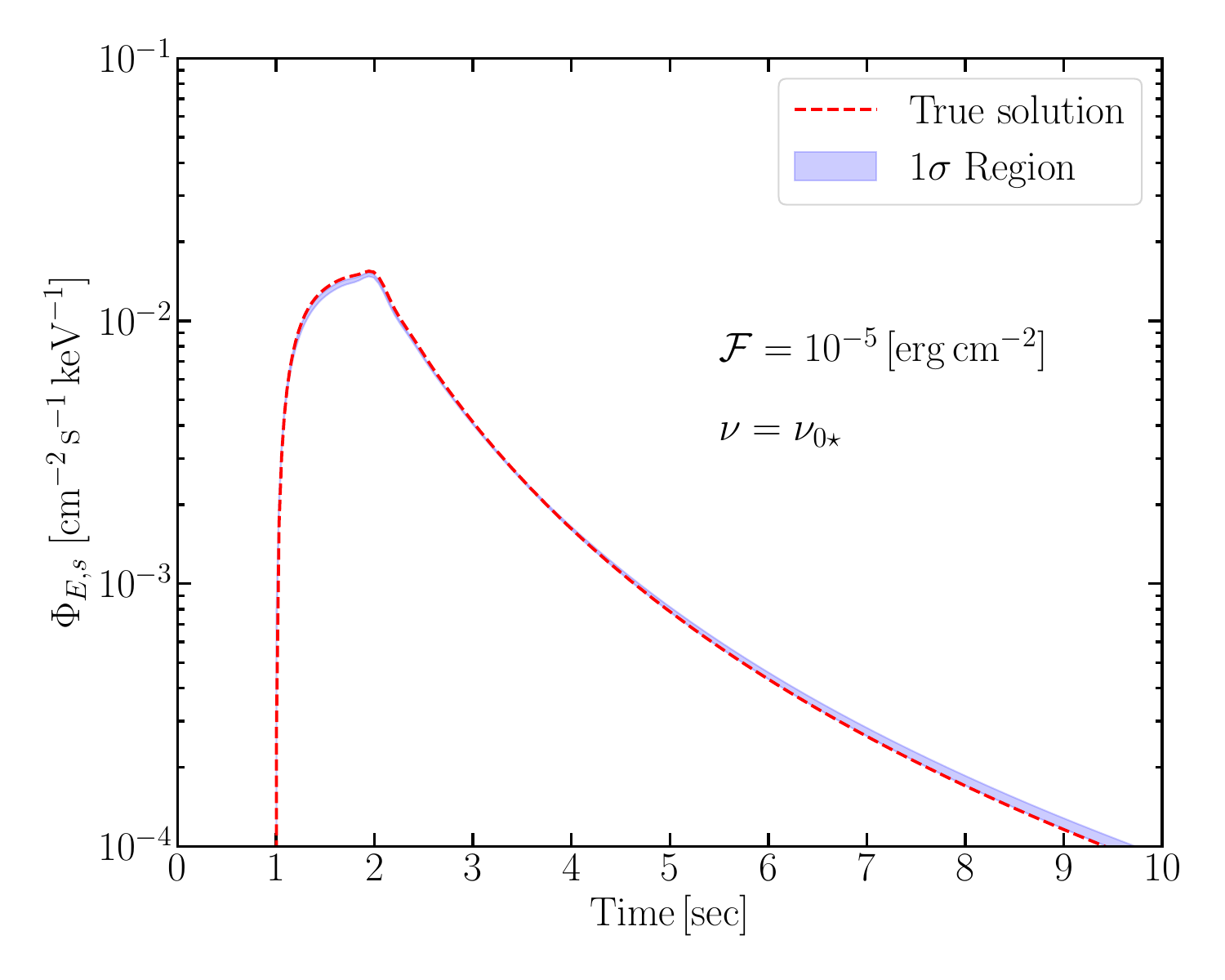}\quad\quad
    \includegraphics[width=0.45\textwidth]{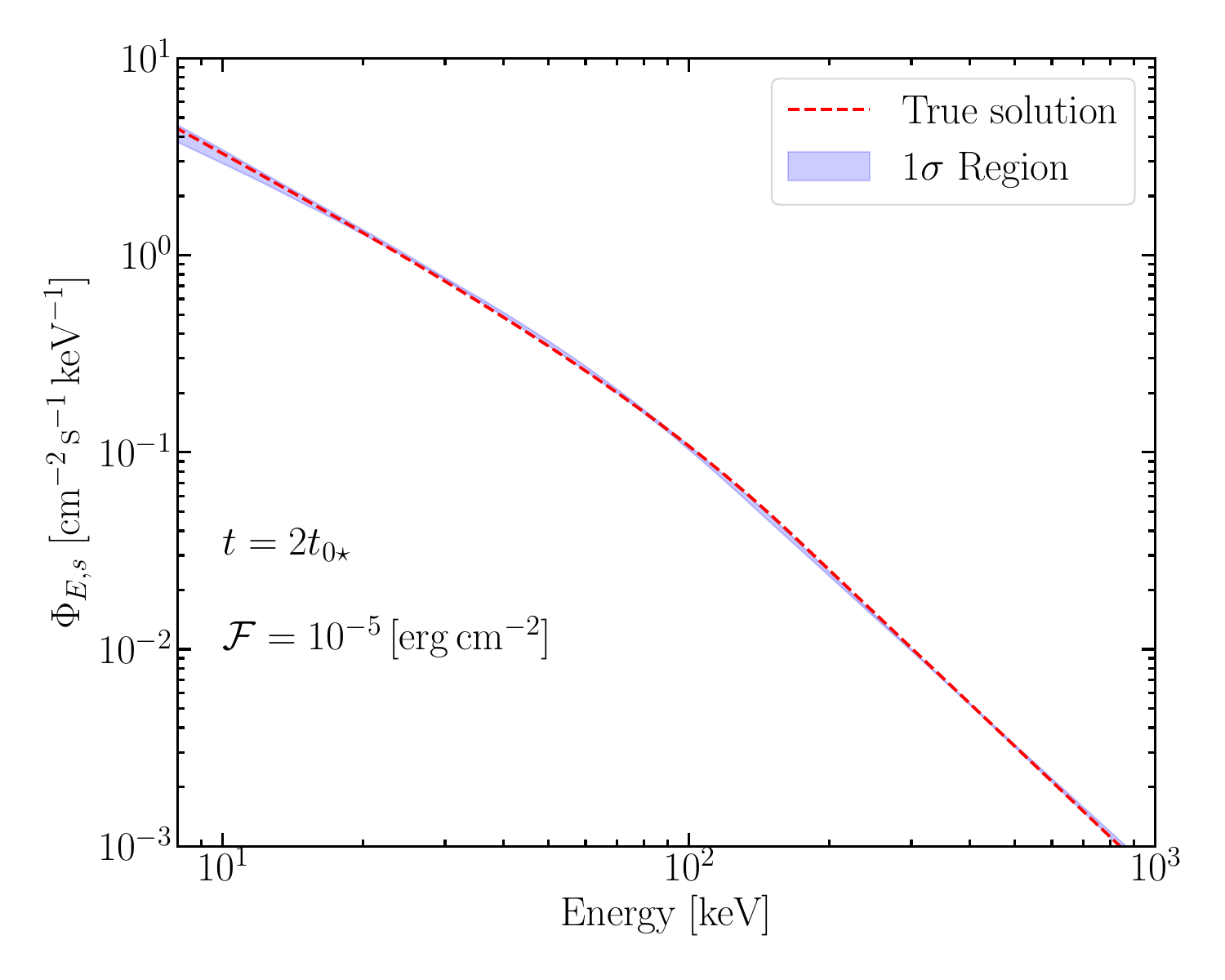}
    \caption{(\textbf{Top}) Model parameter posterior distribution from the joint fit to the lightcuve and time-resolved spectrum of a synthetic source with fluence $\mathcal{F}=10^{-5}\rm{erg\,cm}^{-2}$ plus background for pulse duration of $t_{\rm GRB}=10$\,s. A single random realization of the theoretical model is used to constrain the dynamical parameters, $m$ and $\Delta R/R_0$, and spectral parameters, $E_{\rm pk0}$, $b_1$, and $b_2$. The contours enclose 1-$\sigma$ to 3-$\sigma$ uncertainty regions around the best fit value, and the true value is shown with red lines. 
    (\textbf{Bottom}) Comparison of the synthetic (red dashed line) lightcurve (left) and spectrum (right) shown with the $1\sigma$ uncertainty regions sampled from the posterior parameter distribution. 
    }
    \label{fig:LC-Spec-fit-w-bknd-one-realization}
\end{figure*}
%%%%%%% FIGURE %%%%%%%%%%%%%%%%%%%%%%%%%%%%%%%%%%%%%%%%%%%%%

%%%%%%%%%%%%%%%%%%%%%%%%%%%%%%%%%%%%%%%%
\subsection{Polarized Test Source}
%%%%%%%%%%%%%%%%%%%%%%%%%%%%%%%%%%%%%%%%
Once the dynamical and spectral parameters are constrained (IV.\,A) the temporal evolution of the polarization degree (PD) is determined. In general, the polarization curve also has some spread around the best-fit curve, where the spread results from the $1\sigma$ uncertainties in the best-fit model parameters. As demonstrated in section\,\ref{sec:results}, the spread around the best-fit lightcurve and spectrum will be rather narrow for GRBs with fluence $\mathcal{F}\gtrsim10^{-5}\,{\rm erg\,cm}^{-2}$. The same is expected for the polarization curve which can be constrained well for GRBs with this level of fluence. Therefore, for such bright sources it is safe to simply use the best-fit dynamical and spectral model parameters obtained earlier.

To obtain the list of detected Compton events, we first create the detected modulation curves (IV.\,B), $M(\phi,t\vert\theta_{d,\Pi},\mathbf{C},\Theta_{d,s})\equiv M(\phi,t)=dN/d\phi\,dt$, both for a completely polarized source, i.e. with $\Pi=1$ and polarization angle (PA) $\theta_{d,\Pi}=\theta_0+\theta_\Pi$, and a completely unpolarized source. This is done using the source photon spectrum and the detector's Compton response, $R^{\rm C}(\phi,E\vert\theta_{d,\Pi},\Theta_{d,s})=dA^C_{\rm eff}(\phi,E\vert\theta_{d,\Pi},\Theta_{d,s})/d\phi\equiv R^{\rm C}(\phi,E)$, for fully polarized and unpolarized photons,
\begin{eqnarray}
    M^{\rm Pol}(\phi,t) &=&\int dE\,\Phi_{E,s}(E,t) R_{\rm Pol}^{\rm C}(\phi,E)\;, \\
    M^{\rm UnPol}(\phi,t) &=& \int dE\,\Phi_{E,s}(E,t) R_{\rm UnPol}^{\rm C}(\phi,E) \;.
\end{eqnarray}
To obtain the modulation curve for any time-dependent PD in the interval $0\leq\Pi(E,t)\leq1$, we construct a weighted modulation curve
\begin{equation}
    % M(\phi,t,\Pi,\theta_\Pi) = \Pi M^{\rm Pol}(\phi,t,\theta_\Pi) + (1-\Pi) M^{\rm UnPol}(\phi,t,\theta_\Pi)\,.
    M(\phi,t,\Pi) = \Pi M^{\rm Pol}(\phi,t) + (1-\Pi) M^{\rm UnPol}(\phi,t)\,.
\end{equation}
Here the temporal evolution of the PD is obtained from the theoretical model, \confirm{as shown in panel (e) of Fig.\,\ref{Fig:lc-spec-fit-demo}}, that depends on the dynamical and spectral model parameters, which were obtained earlier. For the polarization background we only consider the CXB that dominates the energy range of the detector. Since it is expected to be unpolarized, we construct the background modulation curve from
\begin{equation}
    M_{\rm bkg}(\phi,t) = \int dE\,\Phi_{E,\rm bkg}(E,t) R_{\rm UnPol}^{\rm C}(\phi,E) \,,
\end{equation}
where the background photon spectrum as a function of the incident photon energy is shown in the left panel of Fig.\,\ref{fig:backgorund_input}. Again, we assume for simplicity that the background is constant in time and produces a constant modulation curve. The total number of detected events can now be obtained by integrating over $\phi$ and $t$, so that
\begin{equation}
    N_{C,s}^{\rm det} + N_{C,\rm bkg}^{\rm det} = \int d\phi\,dt M(\phi,t,\Pi) + \int d\phi\,dt M_{\rm bkg}(\phi,t)\,.
\end{equation}

Next, we obtain the list of source plus background events comprising a scattering azimuth and detection time, $\{\phi,t\}$, from their respective modulation curves using MC sampling (IV.\,C) as done before. The distribution of source events thus obtained (IV.\,D) is shown in panel \confirm{(f)} of Fig.\,\ref{Fig:lc-spec-fit-demo}. Notice that in this figure the scattering angle distribution is shown to span over $0\leq\phi\leq\pi$ whereas it spans over $0\leq\phi\leq2\pi$ in the Compton response shown in Fig.\,\ref{fig:eff_area}. Since the response is cyclic over an angular period of $\phi=\pi$, we have simply mapped the events with $\pi\leq\phi\leq2\pi$ to the interval shown here. Since the Compton effective area declines sharply below 50\,keV, where most of the incident photons are found, the number of detected Compton events is drastically reduced as compared to the total number of detected photons. This demonstrates the need for sources with high fluence to enhance the 
%SNR 
signal to noise ratio for a robust measurement of polarization.

The probability distribution for detecting a Compton event with azimuthal scattering angle $\phi_i$ at time $t_i$ can now be constructed using
\begin{equation}
    P_i^C(\phi_i,t_i|\mathbf C,\Pi,\theta_{d,\Pi},\Theta_{d,s}) = \frac{M(\phi,t,\Pi) + M_{\rm bkg}(\phi,t)}{N_{C,s}^{\rm det} + N_{C,\rm bkg}^{\rm det}}\,.
\end{equation}
This probability distribution is used to construct the log-likelihood, as in Eq.\,(\ref{eq:log-likelihood}), to perform the fit to observations.

%%%%%%%%%%%%%%%%%%%%%%%%%%%%%%%%%%%%%%%%%%%%%%%%%%%%%%%
\section{Results}\label{sec:results}
%%%%%%%%%%%%%%%%%%%%%%%%%%%%%%%%%%%%%%%%%%%%%%%%%%%%%%%

We perform an unbinned model fit to the synthetic \textit{test} source prepared for a fluence of $\mathcal{F}=10^{-5}\,{\rm erg\,cm^{-2}}$. A joint fit over the lightcurve and spectrum is performed first to obtain the dynamical and spectral parameters. Using the best-fit parameters from this fit a polarized test source is prepared, which is then fitted using the same unbinned method. More importantly, we quantify the effect of fluence, ranging from dim to bright bursts, on how well the model parameters can be constrained. For our test case, we fix the model parameters to the following values: $\theta_{d,s}=0$, $\varphi_{d,s}=0$, $\Delta R/R_0=1$, $m=0$, $a=1$, $d=-1$, $b_1=-0.25$, $b_2=-1.25$, $q=0$, $\xi_j=100$, $\nu_0=250\,$keV, $t_0=1\,$s, $t_{\rm GRB} = 10$\,s.

%%%%%%%%%%%%%%%%%%%%%%%%%%%%%%%%%%%%%%%
\subsection{Fit To Lightcurve and Spectrum}
%%%%%%%%%%%%%%%%%%%%%%%%%%%%%%%%%%%%%%%

%%%%%%% FIGURE %%%%%%%%%%%%%%%%%%%%%%%%%%%%%%%%%%%%%%%%%%%%%
\begin{figure}
    \centering
    \includegraphics[width=0.48\textwidth]{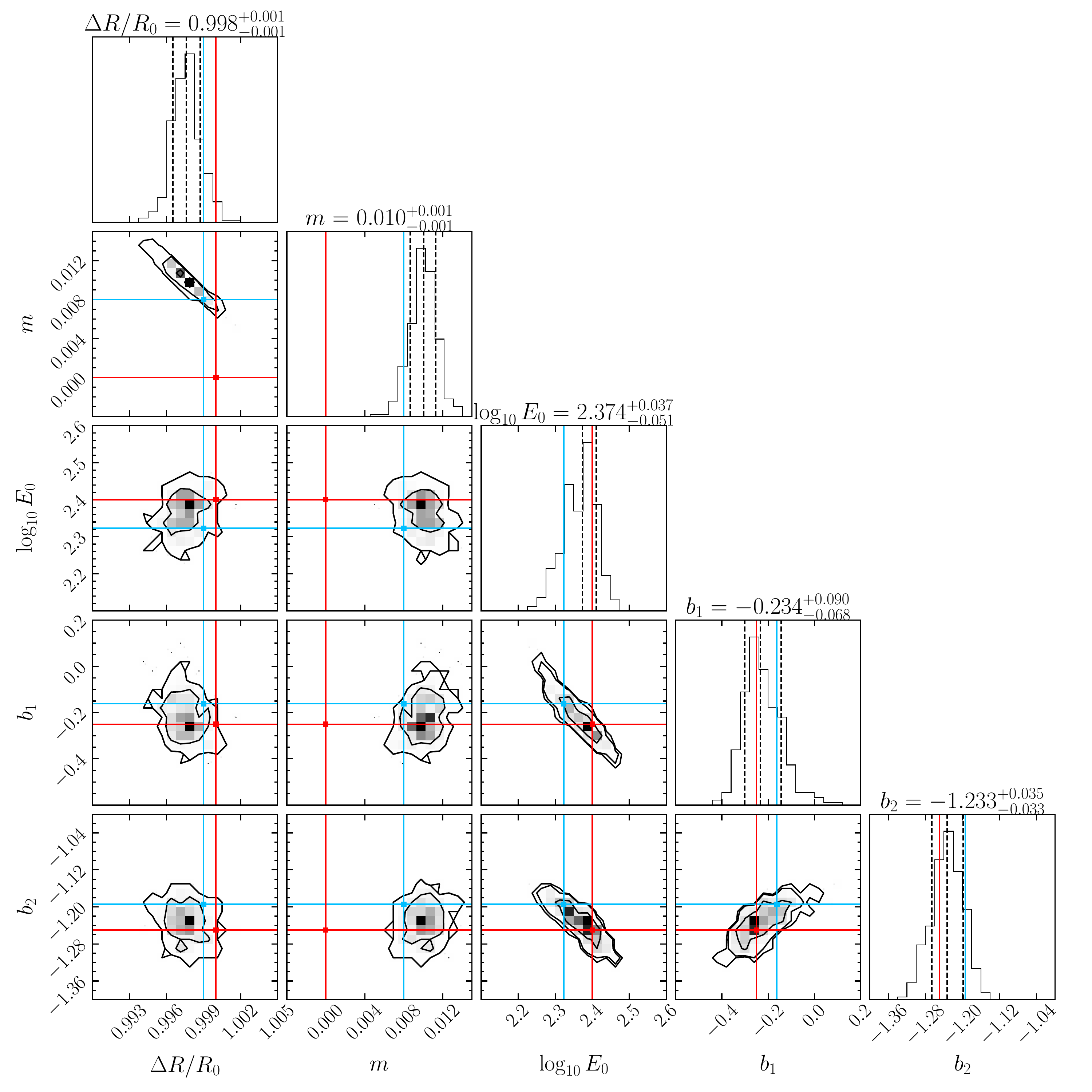}
    \caption{Distributions of the best-fit model parameters obtained from fitting to $300$ random realizations of the theoretical model plus background photons, prepared using the true model parameter values (red lines). The best-fit solution (blue lines) obtained in Fig.\,\ref{fig:LC-Spec-fit-w-bknd-one-realization} for a single random realization of the model lies more than $1\sigma$ away from the true solution for some model parameters.
    }
    \label{fig:LC-Spec-fit-multiple-realizations-from-true-params}
\end{figure}
%%%%%%% FIGURE %%%%%%%%%%%%%%%%%%%%%%%%%%%%%%%%%%%%%%%%%%%%%

%%%%%%% FIGURE %%%%%%%%%%%%%%%%%%%%%%%%%%%%%%%%%%%%%%%%%%%%%
\begin{figure*}
    \centering
    \includegraphics[width=0.33\textwidth]{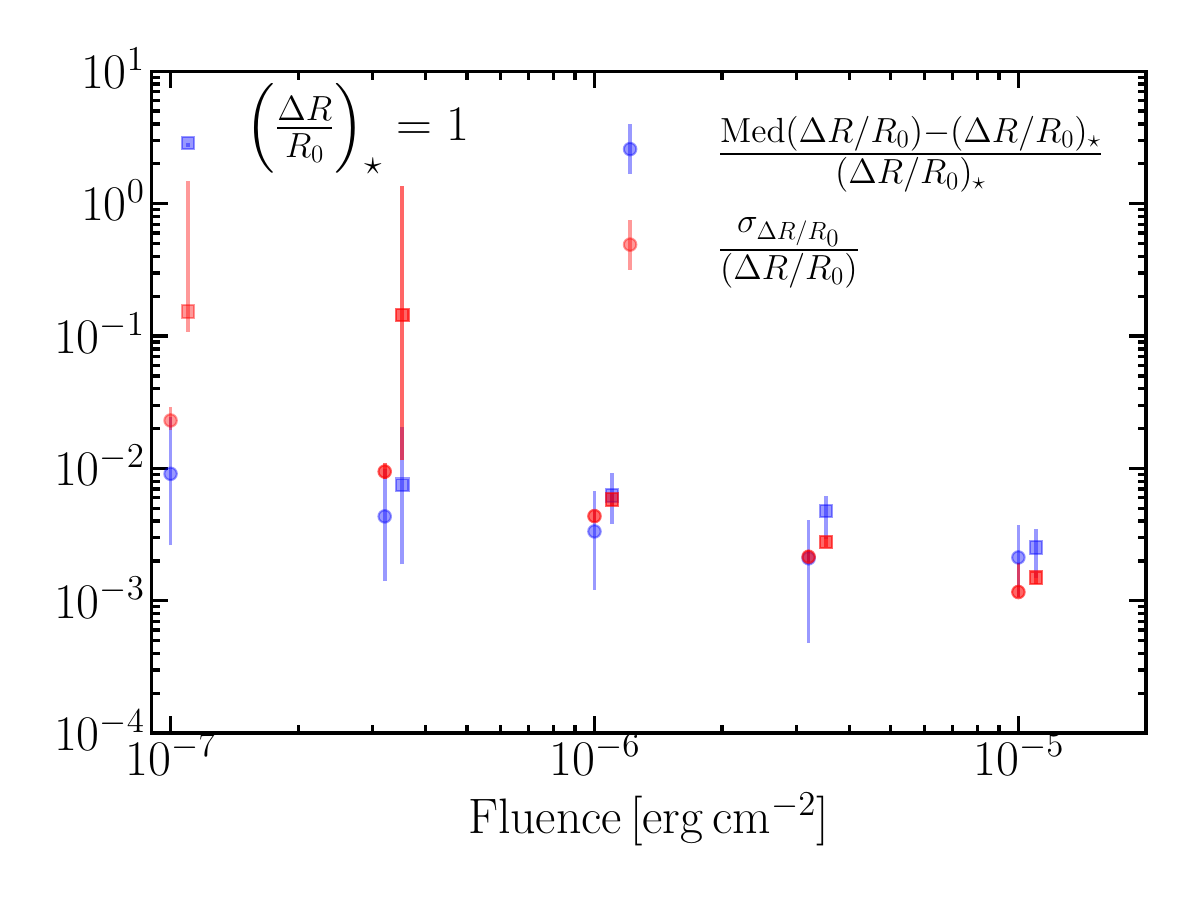}
    \includegraphics[width=0.33\textwidth]{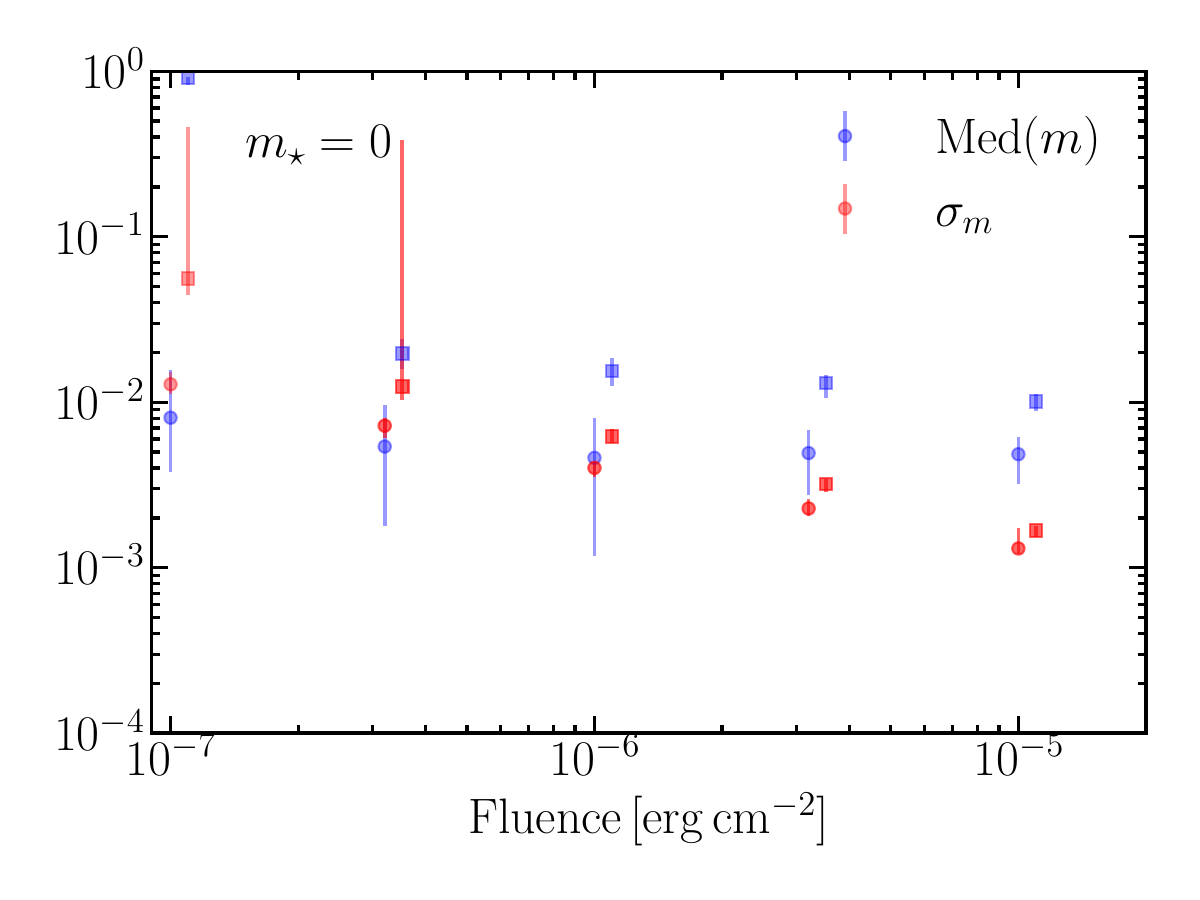} \\
    \includegraphics[width=0.33\textwidth]{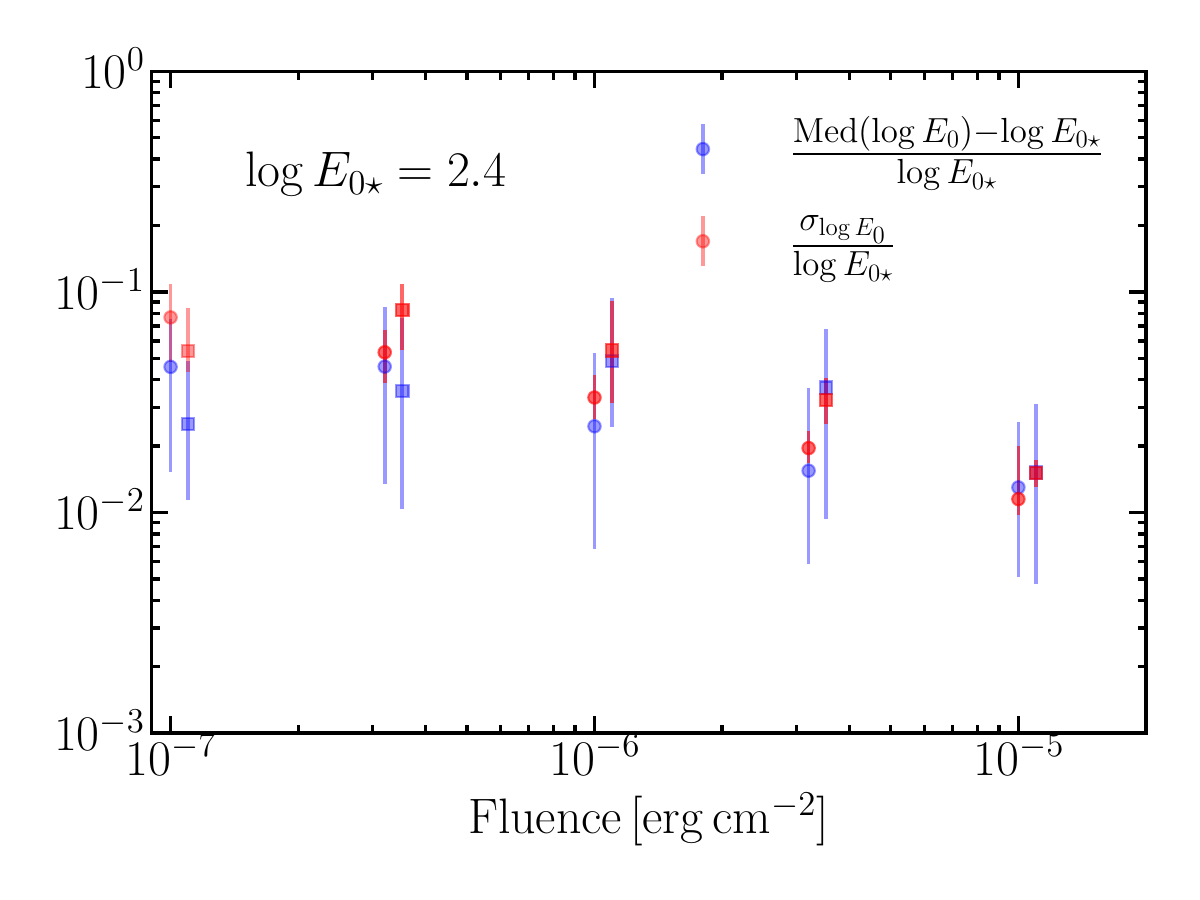}
    \includegraphics[width=0.33\textwidth]{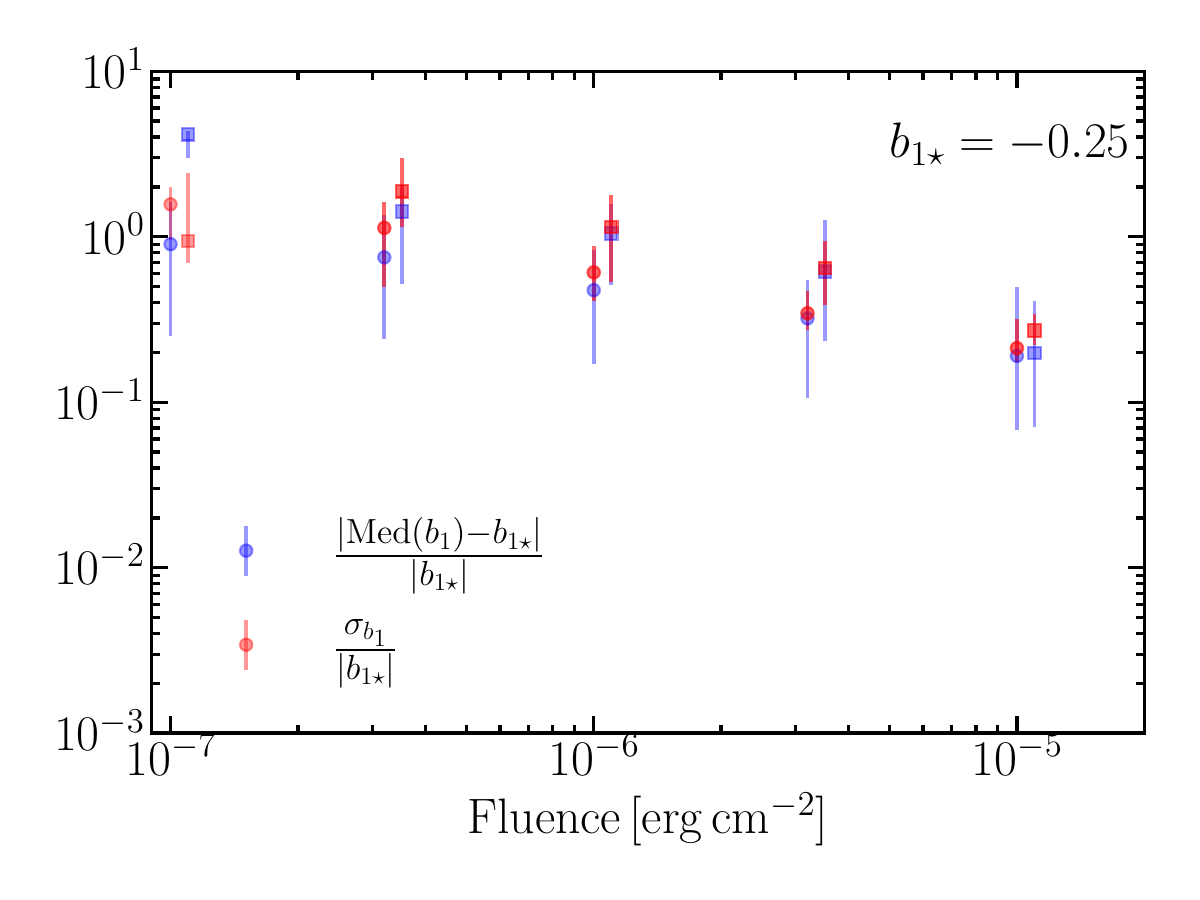}
    \includegraphics[width=0.33\textwidth]{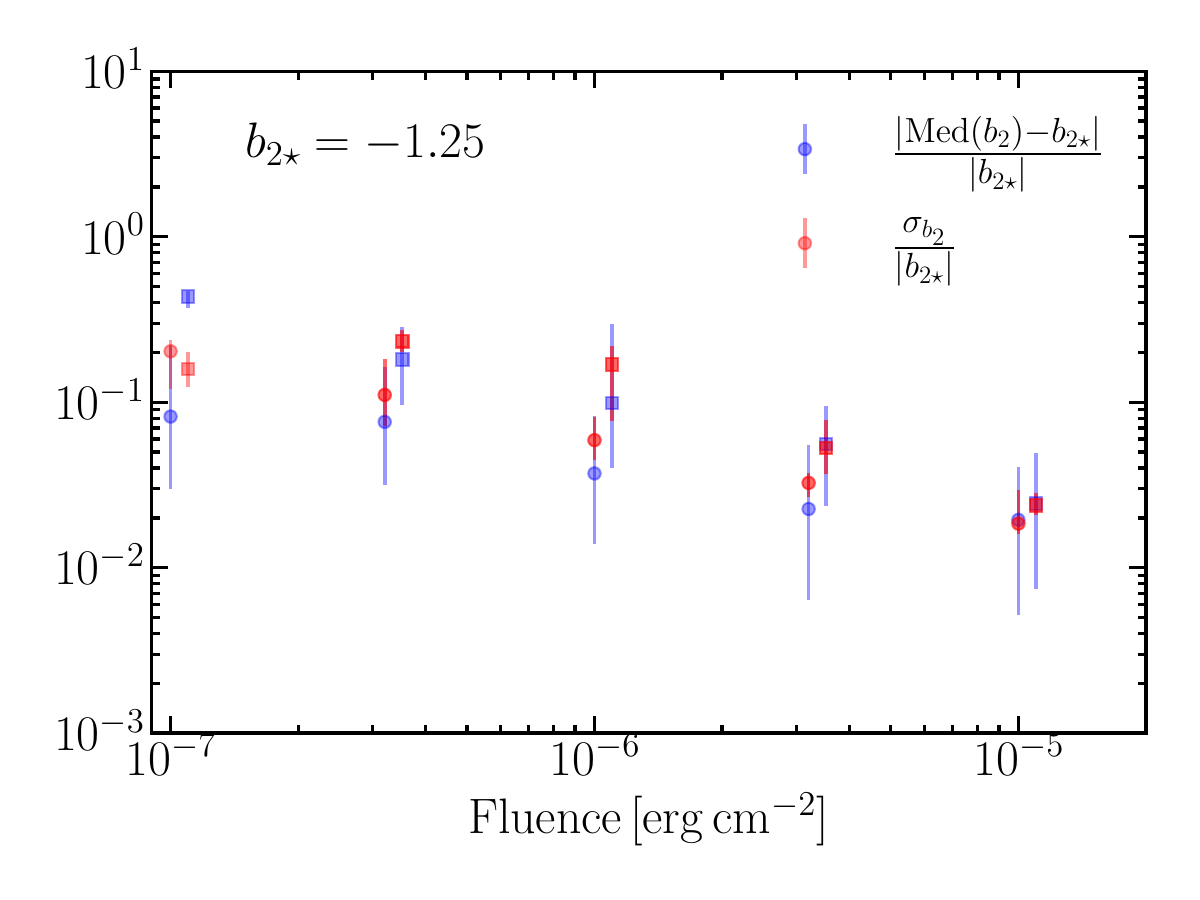}
    \caption{Accuracy and precision of the best-fit model parameters as a function 
    of source fluence, obtained by performing $100$ random realizations of 
    the model prepared using the true values shown with a star in all panels. 
    Accuracy (blue points) is shown by the difference between the true value and median (or best-fit) value 
    of the posterior distribution of the model parameters obtained in each random 
    realization. The precision (red points) is shown by the $1\sigma$ spread of 
    the posterior distribution obtained in each random realization around the best-fit 
    (or median) value. Data shown using circles only include source photons and that 
    shown with squares (artificially shifted in fluence by a multiplicative factor of 1.1 
    to prevent overlapping) include background photons.
    }
    \label{fig:LC-Spec-fit-error-Vs-fluence}    
\end{figure*}
%%%%%%% FIGURE %%%%%%%%%%%%%%%%%%%%%%%%%%%%%%%%%%%%%%%%%%%%%

We start by jointly fitting the theoretical model to both the lightcurve and time-resolved spectrum since they both are coupled. In fitting either of them separately would require fixing either the spectral or dynamical parameters, reducing the accuracy of the best-fit parameters. The fitting procedure follows the methodology demonstrated in Figures\,\ref{fig:flowchart} and \ref{Fig:lc-spec-fit-demo} for creating and fitting a single realization (point II in Fig.\,\ref{fig:flowchart}) of the test source using MCMC.

The top panel of Fig.\,\ref{fig:LC-Spec-fit-w-bknd-one-realization} shows the posterior distributions of the dynamical ($\Delta R/R_0$, $m$) and spectral ($E_0$, $b_1$, $b_2$) model parameters for a single random realization (or sampling) of the theoretical model that was used to prepare the test case. The true values of all the parameters are shown with red lines. A single random realization represents an observation made by a given detector. A similar detector observing the same event would have detected another random sample of photons from the source. The posterior distribution of the model parameters after fitting to the latter random sample would look very similar to that shown in Fig.\,\ref{fig:LC-Spec-fit-w-bknd-one-realization}, but with slightly different best-fit values. The bottom panel of Fig.\,\ref{fig:LC-Spec-fit-w-bknd-one-realization} shows the lightcurve and spectrum prepared using the true solution (dashed red curves) as well as the $1\sigma$ spread obtained from randomly sampling the model parameter posterior distributions.

Both the posterior distributions and the $1\sigma$ spread in the lightcurve and spectrum show a narrow distribution of solutions around the best-fit solution. The narrowness signifies the precision but not the accuracy, where the latter can be judged from the separation between the red and blue lines in the marginalized parameter distributions. The contours show that the true values of all of the parameters, except for $m$, lie within $1\sigma$ of the best fit value (blue lines). The best-fit solution does not coincide with the true solution due to the presence of systematic and random errors, where the former stems from the fitting procedure and the latter from finite sampling of the source spectrum and lightcurve. The systematic error will remain even in the absence of any background signal.

The magnitude of both errors can be quantified by constraining one of the two. The systematic error can be understood by fitting extremely high fluence sources, in which case the random error will be diminished by having very large number of source photons. Of course, the task will become computationally expensive and even impractical. On the other hand, the random error can be reduced, in theory, by making several measurements of the same event. This is not possible for transient and variable sources in practice. However, in the present case, since we know the exact theoretical model, the random error can be reduced by fitting the model to several (or a large number) random realizations (or observations) of the theoretical model (or source) prepared with the same model parameters. The underlying expectation being that the peak of the distribution of a large number of best fit values will be much closer to the true value of the fitted parameter. 

This is demonstrated in Fig.\,\ref{fig:LC-Spec-fit-multiple-realizations-from-true-params} where we first prepare $300$ random realizations of the theoretical model and then carry out the MCMC fits. This figure shows the distribution of the best-fit values for all the model parameters obtained in the $300$ different fits of the same source. Indeed, the peak of the distributions of the spectral parameters is now much closer to the true values. The same is not true for the dynamical parameters that show a larger difference between the distribution peak and the true solution. This residual error is now dominated by the systematic error, which we deal with below. This figure also shows the location of the best-fit parameters obtained in a single fit to the synthetic data in Fig.\,\ref{fig:LC-Spec-fit-w-bknd-one-realization} (blue line). In general, we are expected to obtain only one set of model parameters out of the distribution shown in Fig.\,\ref{fig:LC-Spec-fit-multiple-realizations-from-true-params}. Depending on how narrow or wide the distributions are in this figure, which is sensitive to the source fluence, we may obtain a best-fit solution which is either farther or closer, respectively, to the true value.

%%%%%%%%%%%%%%%%%%%%%%%%%%%%%%%%%%%%%%%%%%%%%%%%%%%%%%%%%%%%%%%%%%
\subsubsection{Model Parameter Uncertainties Vs Source Fluence}
%%%%%%%%%%%%%%%%%%%%%%%%%%%%%%%%%%%%%%%%%%%%%%%%%%%%%%%%%%%%%%%%%%
The level of fluence of the GRB prompt emission makes a significant difference in the accuracy of best fit model parameters. Most GRBs record a fluence of $\mathcal{F}\lesssim10^{-5}\,{\rm erg\,cm}^{-2}$, the limit below which many GRBs may be too photon starved to obtain statistically significant polarization measurements (see below). Furthermore, at low fluence levels for fixed GRB duration or for longer duration GRBs at a fixed fluence, the effect of the instrument background may start to become significant. Since the background photons continue to accumulate over the GRB duration, the signal to noise ratio is reduced in GRBs with longer lasting emission episodes for the same fluence. Therefore, to assess how the level of fluence as well as instrumental background affects the quality of the best-fit model parameters, we perform fits to 100 random realizations for different values of the source fluence. Figure\,\ref{fig:LC-Spec-fit-error-Vs-fluence} shows the accuracy of the best-fit parameters when compared with the true solutions (blue points) as well as the precision, i.e. the $1\sigma$ errors around the best-fit value (red points), as a function of fluence. We show results with (filled squares) and without (filled circles) background photons, to assess the level of fluence below which the results become background dominated. 

In general, and as expected, model parameter estimation becomes slightly poor when background photons are included. Only at $\mathcal{F}\gtrsim10^{-5}\,{\rm erg\,cm^{-2}}$, and for $t_{\rm GRB}=10$\,s, the effect of the background diminishes to the extent that the results start to become predominantly source dominated. Inference of model parameters becomes particularly worse for $\mathcal{F}\lesssim10^{-6}\,{\rm erg\,cm^{-2}}$ when the $1\sigma$ uncertainties in the best-fit solutions starts to exceed the accuracy of the parameters. Above this fluence the error scales approximately as $(N_{\rm ph}^{\rm det})^{-1/2} \propto \mathcal{F}^{-1/2}$, which is expected due to Poisson noise. Therefore, to obtain errors as small as a few per cent on the dynamical and spectral parameters in our model from the lightcurve and spectral analysis, GRBs with fluence exceeding $10^{-5}\,{\rm erg\,cm^{-2}}$ will be needed.

%%%%%%%%%%%%%%%%%%%%%%%%%%%%%%%%%%%%%%%%%%%%%%%%%%%%%%%%%%%%%%%%%%
\subsubsection{Removal of Systematic Bias}
%%%%%%%%%%%%%%%%%%%%%%%%%%%%%%%%%%%%%%%%%%%%%%%%%%%%%%%%%%%%%%%%%%
Next, we attempt to remove the systematic bias from the best-fit solution obtained in a single realization by preparing a new set of $N=300$ random realizations (point III from Fig.\,\ref{fig:flowchart}), but this time using the best-fit model parameter values as obtained in Fig.\,\ref{fig:LC-Spec-fit-w-bknd-one-realization}. This situation is similar to what is expected in practice. We will observe a given GRB with POLAR-2 and proceed to fit the data with our theoretical model. This will yield one best-fit solution, $\mathbf{C}_0=\{A_0, B_0, ...,\}$, out of the distribution around the true values of the model parameters, $\mathbf{C}_\star=\{A_\star, B_\star,...,\}$. By performing model fits on $N\gg1$ random realizations that are prepared using the best-fit solution $\mathbf{C}_0$, we will reproduce the systematic bias that may have affected the best-fit solution itself.

\begin{figure*}
    \centering
    \includegraphics[width=0.48\textwidth]{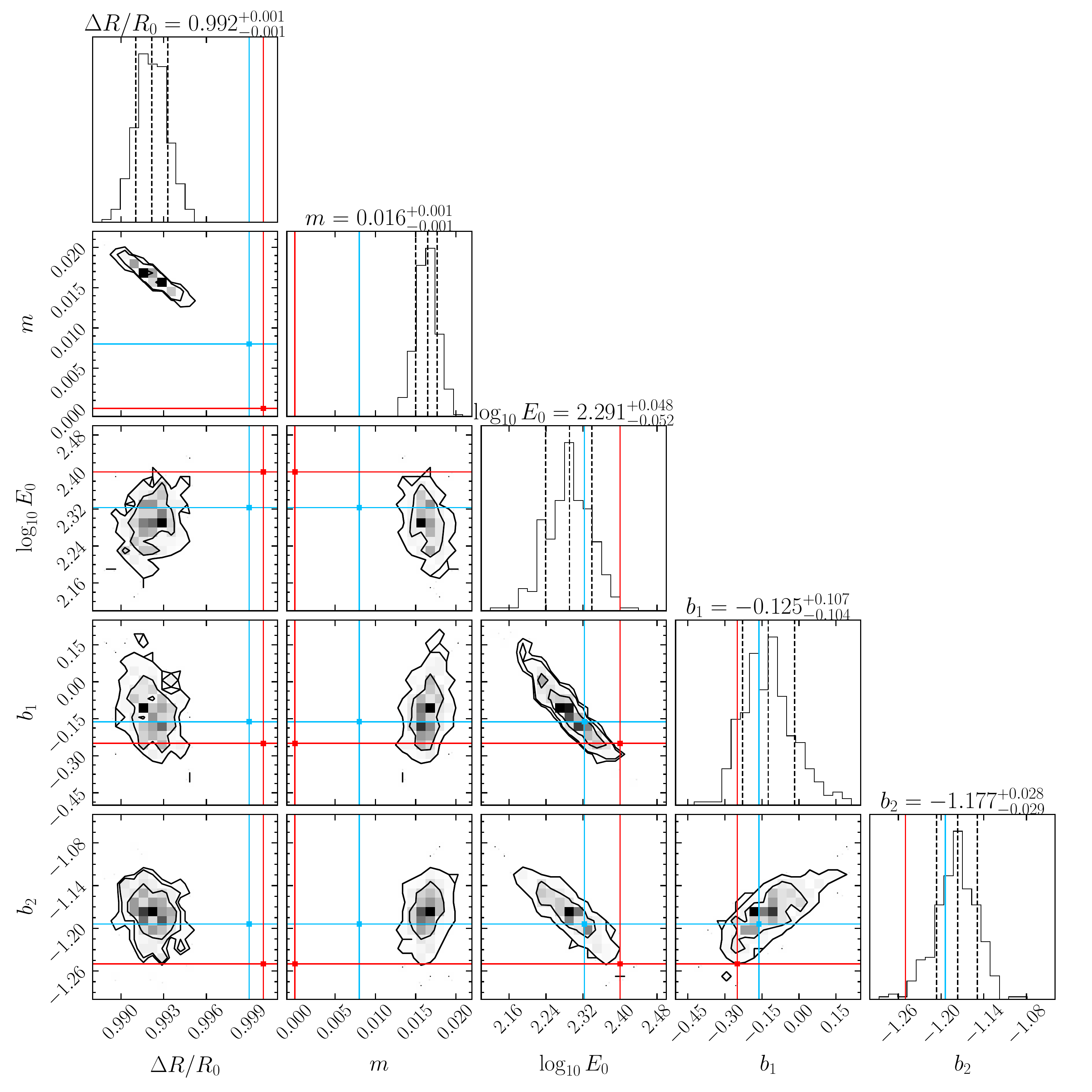}
    \includegraphics[width=0.48\textwidth]{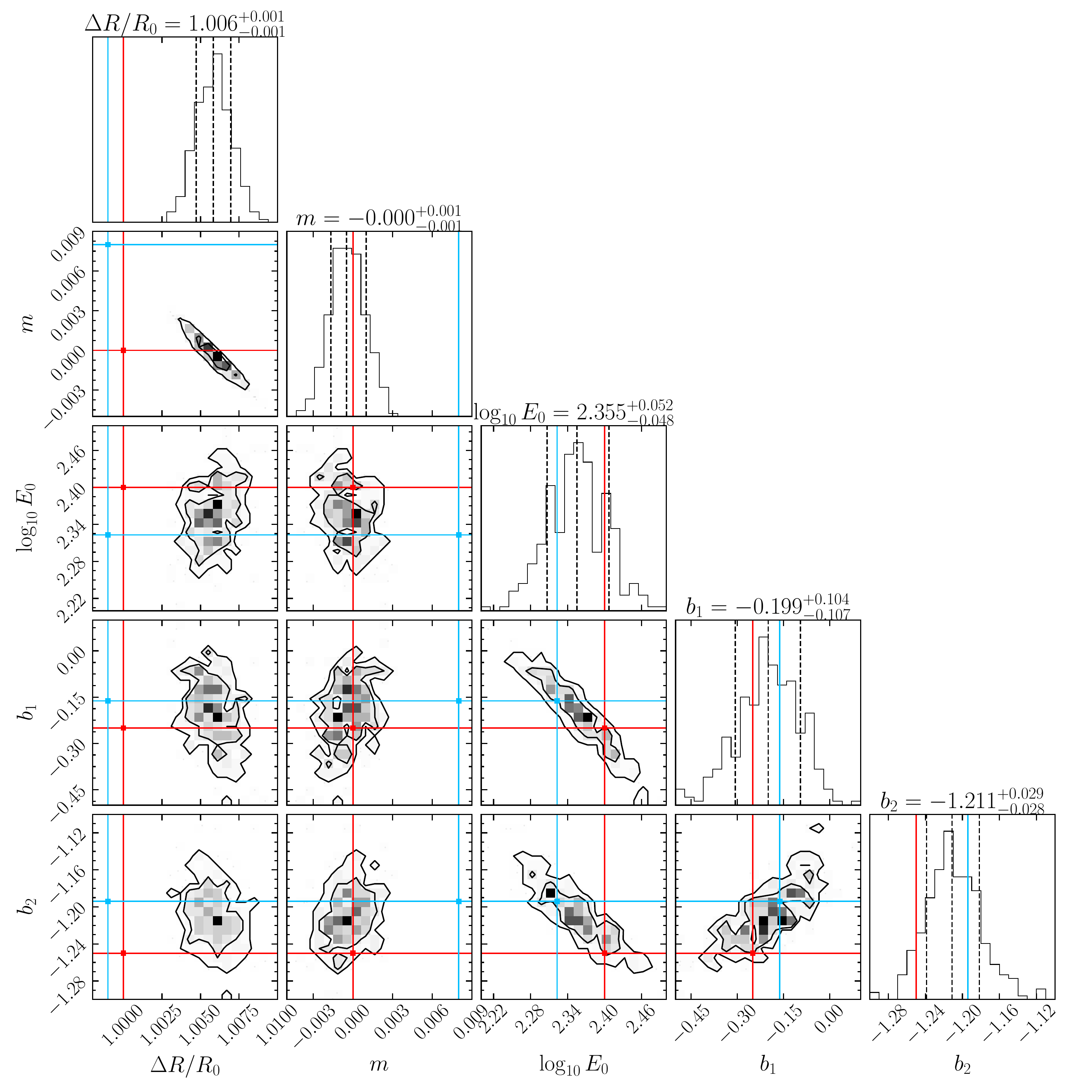}
    \caption{(Left) Distributions of the best-fit model parameters obtained from fitting to $300$ random realizations of the theoretical model prepared using the best-fit model parameter values (blue lines) obtained in Fig.\,\ref{fig:LC-Spec-fit-w-bknd-one-realization}. 
    (Right) Inferred distributions of the true model parameters obtained from inverting (i.e. by removing the systematic bias) the distributions in the left panel. The red lines show the true model parameter values.
    }
    \label{fig:LC-Spec-fit-multiple-realizations}
\end{figure*}

The left panel of Fig.\,\ref{fig:LC-Spec-fit-multiple-realizations} shows the distributions of the best-fit model parameters obtained from fitting $N=300$ random realizations prepared using $\mathbf{C}_0$ (indicated by the blue lines). The peaks of marginalized (one-dimensional) distributions do not align exactly with the best-fit values (blue lines; $\mathbf{C}_0$) and reveal the bias. We remove this bias using the following procedure: The distribution of any two model parameters, e.g. $A$ and $B$, in the two-dimensional space, as shown in the left panel of Fig.\,\ref{fig:LC-Spec-fit-multiple-realizations}, has coordinates $(A_i, B_i)$ for $0\leq i \leq N$ with $N$ being the number of random realizations. This distribution is offset from the expected one $(A_0,B_0)$ by an amount $(\delta A_i, \delta B_i) = (A_i-A_0, B_i-B_0)$. To remove this offset (or bias) and obtain our best unbiased probability distribution for the true model parameters, $\mathbf{C}_\star$, we perform an inversion such that $(A_{\star,i}, B_{\star,i}) = (A_0-\delta A_i, B_0-\delta B_i)$. This procedure is carried out for the other pairs of model parameters to obtain the inverted (i.e. unbiased) distributions, as shown in the right panel of Fig.\,\ref{fig:LC-Spec-fit-multiple-realizations}, which are now closer to the true values.

%%%%%%% FIGURE %%%%%%%%%%%%%%%%%%%%%%%%%%%%%%%%%%%%%%%%%%%%%
%%%%%%% FIGURE %%%%%%%%%%%%%%%%%%%%%%%%%%%%%%%%%%%%%%%%%%%%%

%%%%%%% FIGURE %%%%%%%%%%%%%%%%%%%%%%%%%%%%%%%%%%%%%%%%%%%%%
\begin{figure*}
    \centering
    \includegraphics[width=0.45\textwidth]{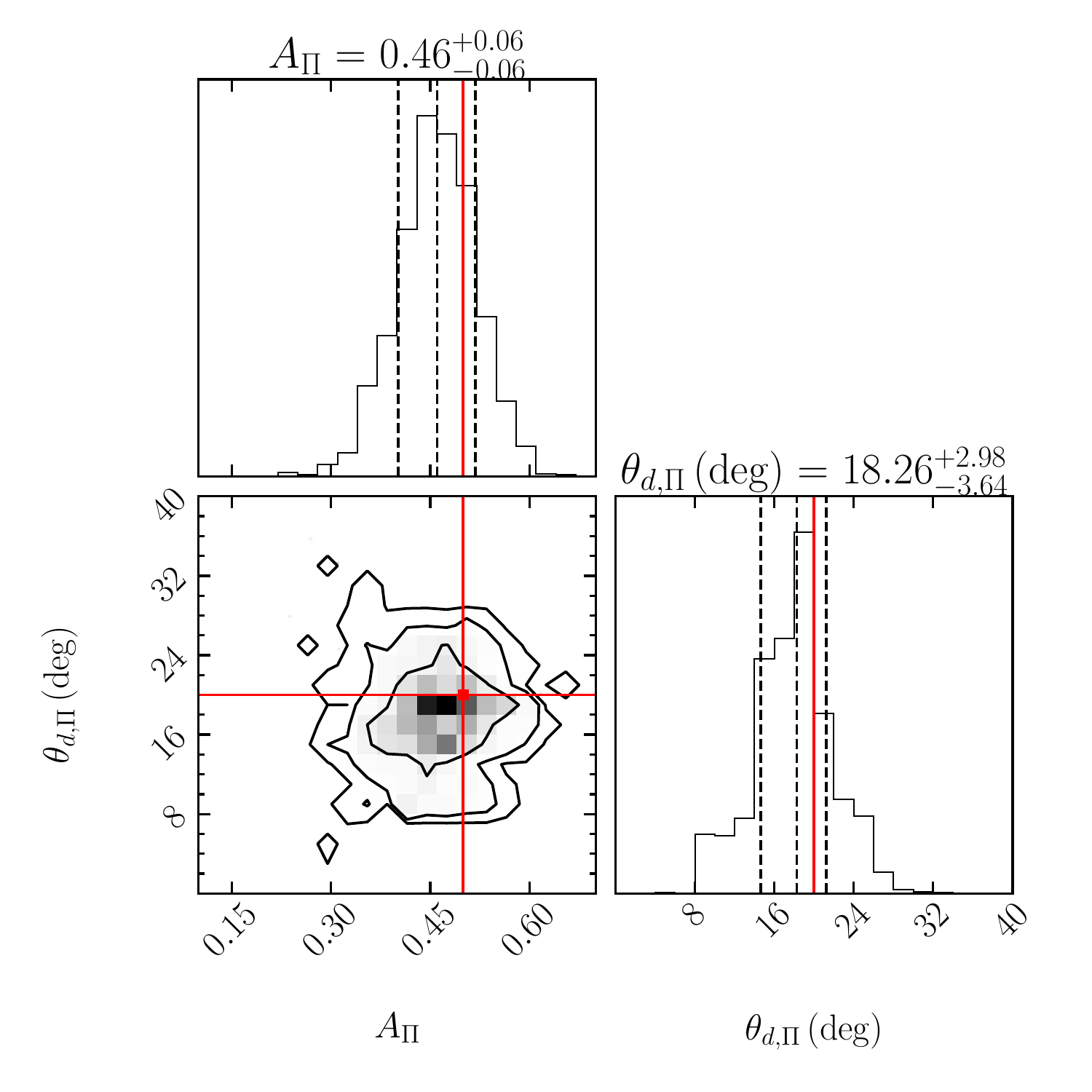}
    \includegraphics[width=0.48\textwidth]{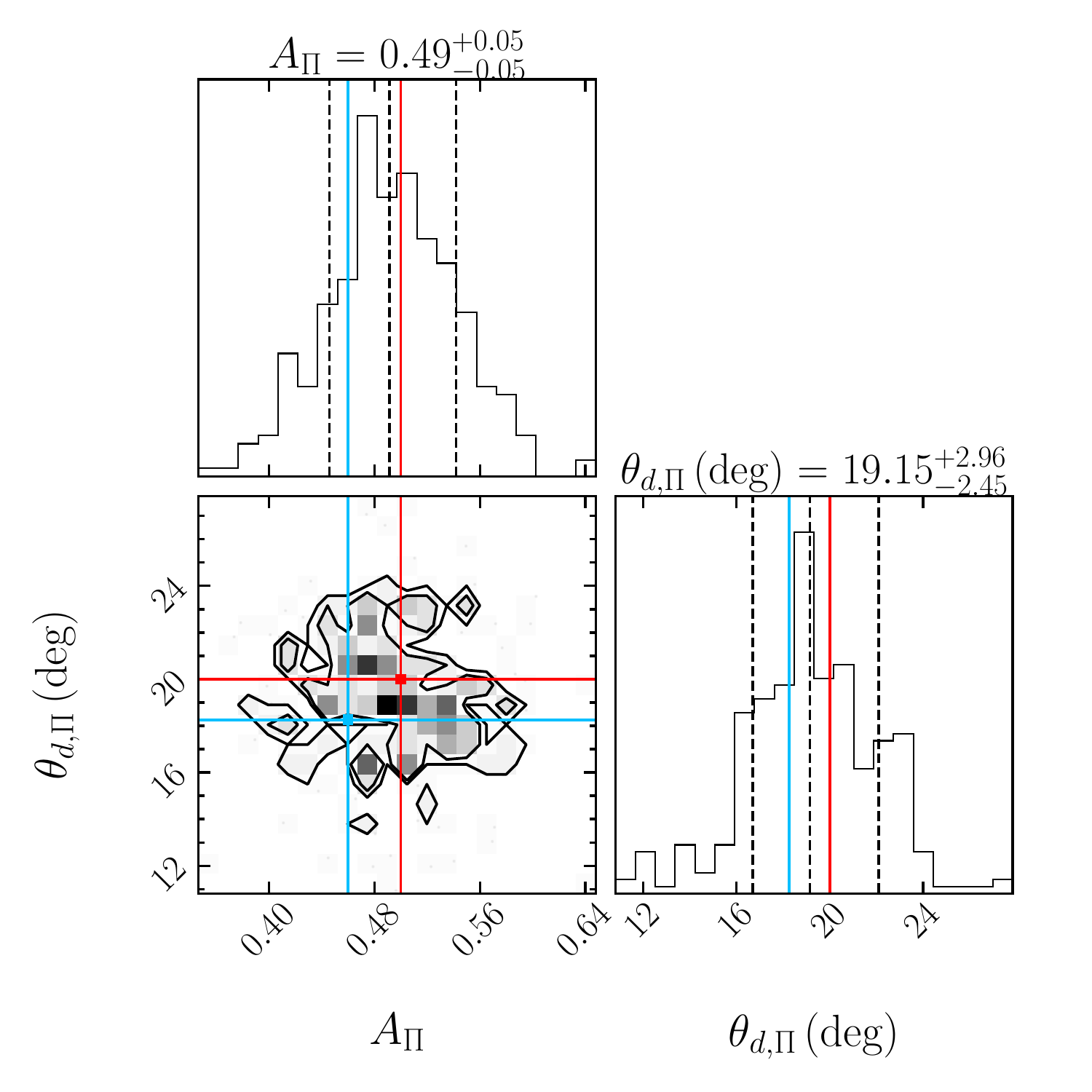} \\
    \includegraphics[width=0.55\textwidth]{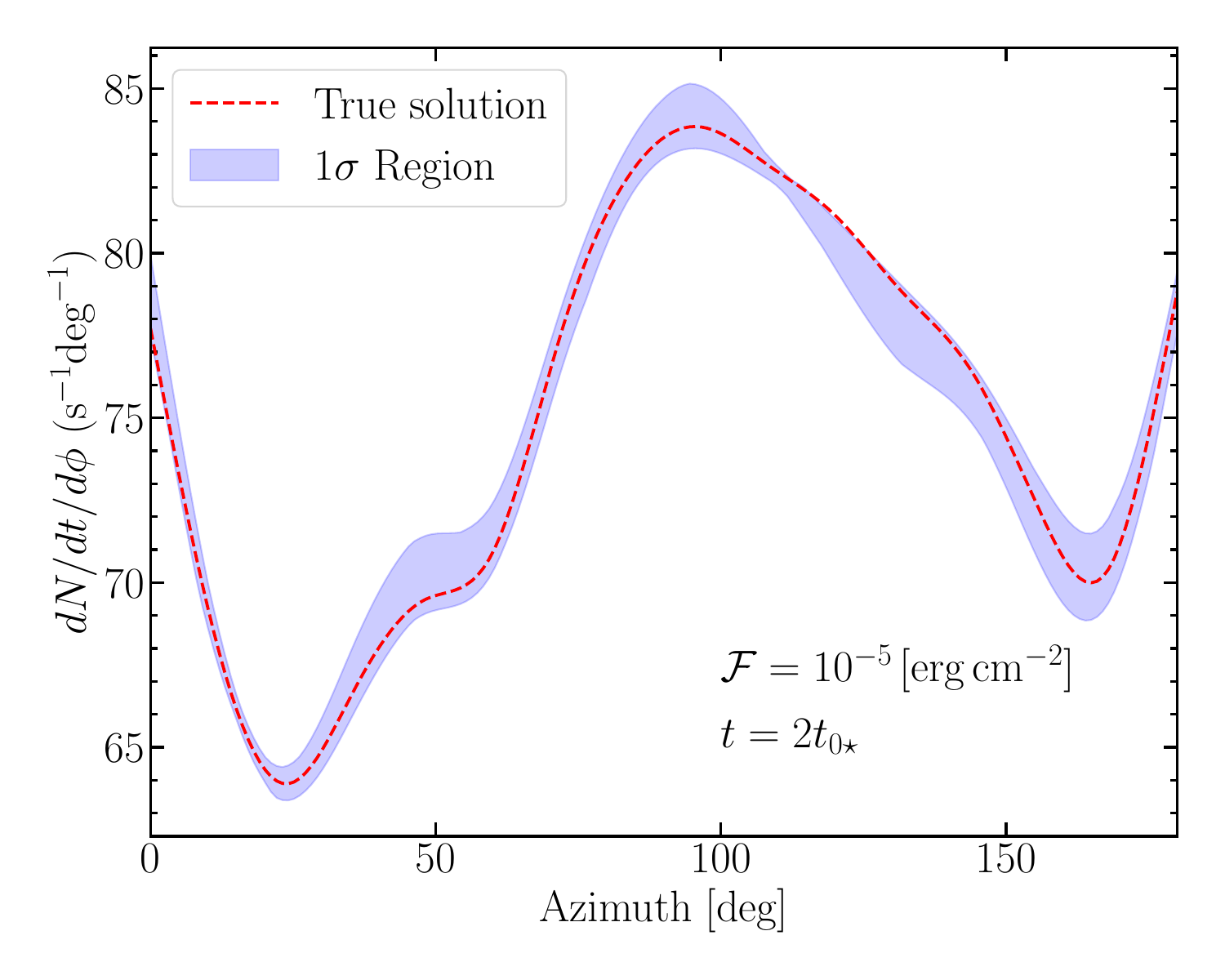}
    \caption{(\textbf{Top}) The posterior distributions for $A_\Pi$ and polarization angle $\theta_{d,\Pi}$ for source fluence $\mathcal{F}=10^{-5}\,{\rm erg\,cm}^{-2}$ with background photons. These are obtained by fitting model predictions to time-resolved polarization degree with a constant polarization angle. Furthermore, the fit was obtained after fixing both the spectral and dynamical model parameters to their best-fit values obtained in a separate fit earlier. 
    (Left) Posterior distributions from a single random realization, and (Right) distributions of the best-fit parameters from $N=300$ random realizations. The red lines show the 
    true model parameter values used to create a synthetic source and the blue lines show the best-fit parameters found in the left panel. 
    (\textbf{Bottom}) $1\sigma$ (blue shaded) region showing the modulation curve sampled from posterior distributions of $A_\Pi$ and $\theta_{d,\Pi}$. The modulation curve obtained from the true parameter values is shown using a red dashed curve.
    }
    \label{fig:A-and-PA-dist-with-bknd-photons}
\end{figure*}
%%%%%%% FIGURE %%%%%%%%%%%%%%%%%%%%%%%%%%%%%%%%%%%%%%%%%%%%%

%%%%%%%%%%%%%%%%%%%%%%%%%%%%%%%%%%%%
\subsection{Fit to Time-Dependent Polarized Emission}
%%%%%%%%%%%%%%%%%%%%%%%%%%%%%%%%%%%%
The three different detectors on POLAR-2 will allow for a robust measurement of the spectrum and its temporal evolution. As a result, the spectral and dynamical parameters of the model can be ascertained first, as shown above, and then used subsequently with polarization measurements to determine the nature of the B-field, viewing geometry ($q=\theta_{\rm obs}/\theta_j$), and even the bulk-$\Gamma$ from $\xi_j=(\Gamma\theta_j)^2$ and $\theta_j$, where the latter can be constrained from jet breaks in the afterglow lightcurve. Thus avoiding a more computationally expensive joint fit over the lightcurve, spectrum, and polarization.

When preparing the polarized test case, we make the simplifying assumption that the B-field is ordered on large scales ($B_{\rm ord}$) and the observer's LOS as well as the beaming cone are well within the jet aperture, so that the emission can be treated as arising from a spherical flow. Under these assumptions, the model parameters $q$ and $\xi_j$ cannot be constrained. In addition, it also restricts the nature of the B-field since only a large-scale ordered B-field can yield non-vanishing polarization, while any shock-produced small scale fields ($B_\perp$ and $B_\parallel$) would produce net zero polarization for this viewing geometry. The different B-field configurations and jet angular structures will be explored in a separate follow-up work.

Once the spectral and dynamical parameters have been determined, the temporal evolution of polarization from a pure $B_{\rm ord}$ field is well defined (see, e.g., Fig.\,3 of \citealt{Gill-Granot-21}), leaving no free parameters, except for the PA, that can be further constrained with polarization. However, in the presence of small-scale random B-fields along with an ordered field component (see, e.g., \citealt{Granot-Konigl-03}, where this case is discussed for afterglow polarization), the degree of polarization will be diluted. In general, the small-scale B-fields can contribute non-negligible polarization for off-axis observers when their LOSs are close to the edge of the jet, e.g., in a top-hat jet. However, for a spherical flow, the net polarization from such small-scale fields vanishes due to symmetry. Nonetheless, they still contribute a fraction of the total intensity, which would dilute the polarization from a purely ordered field. This situation can be parameterized where the total polarization is given by
\begin{equation}
    \Pi(t) = A\Pi_{B_{\rm ord}}(t)\,,
\end{equation}
where $0\leq A\leq 1$ is the dilution factor, which can now be constrained from time-resolved polarization measurements. The dilution factor can potentially have a time-dependence, but here we simply keep it constant. Here $\Pi_{B_{\rm ord}}(t)$ is obtained from the theoretical model for the given set of pre-determined dynamical and spectral parameters. When the field is purely ordered with no small-scale random component, then the dilution factor, $A = \Pi(t)/\Pi_{B_{\rm ord}(t)}$, yields the level of agreement with the theoretical expectations. For example, $A\sim1$ would mean close agreement and $A\ll1$ or $A\gg1$ would mean poor or no agreement. This can be done for other B-field configurations as well, e.g. $B_\perp$ and $B_\parallel$, but in this case the jet must be viewed off-axis to obtain non-zero net polarization. It is also possible to use other goodness of fit measures when comparing time and energy resolved measurements of polarization with theoretical models. These will be explored in a future work.

%%%%%%% FIGURE %%%%%%%%%%%%%%%%%%%%%%%%%%%%%%%%%%%%%%%%%%%%%
\begin{figure*}
    \centering
    \includegraphics[width=0.48\textwidth]{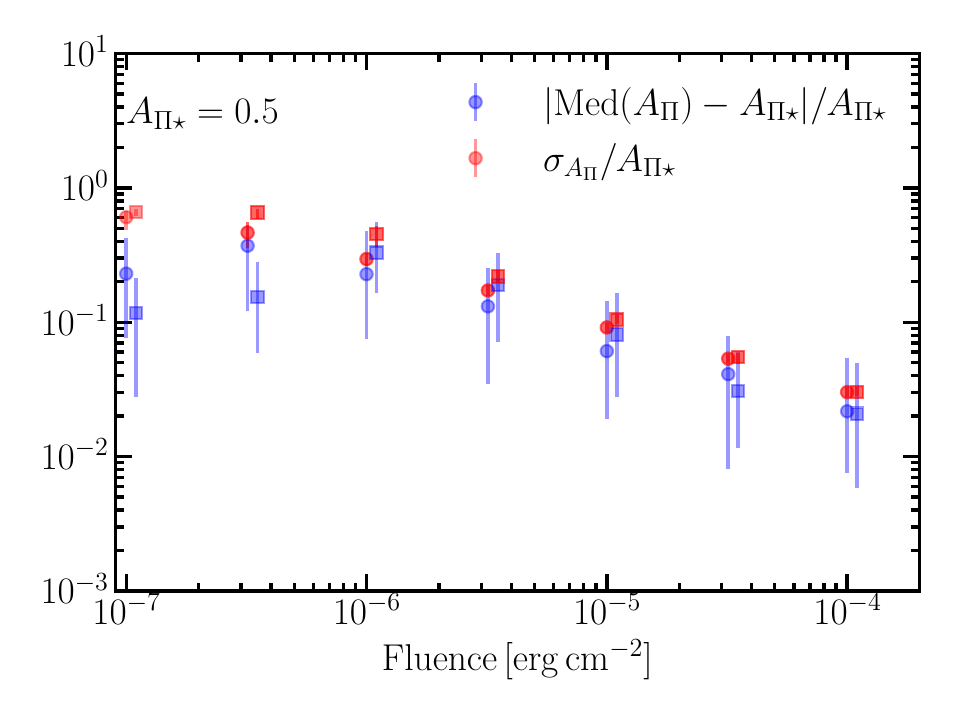}
    \includegraphics[width=0.48\textwidth]{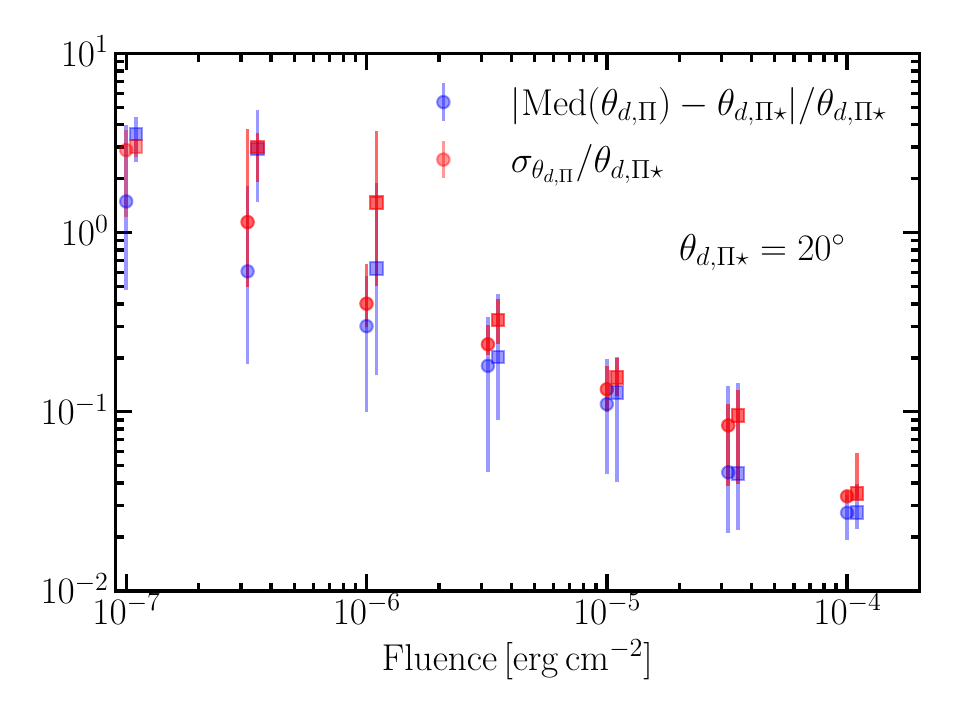}
    \includegraphics[width=0.48\textwidth]{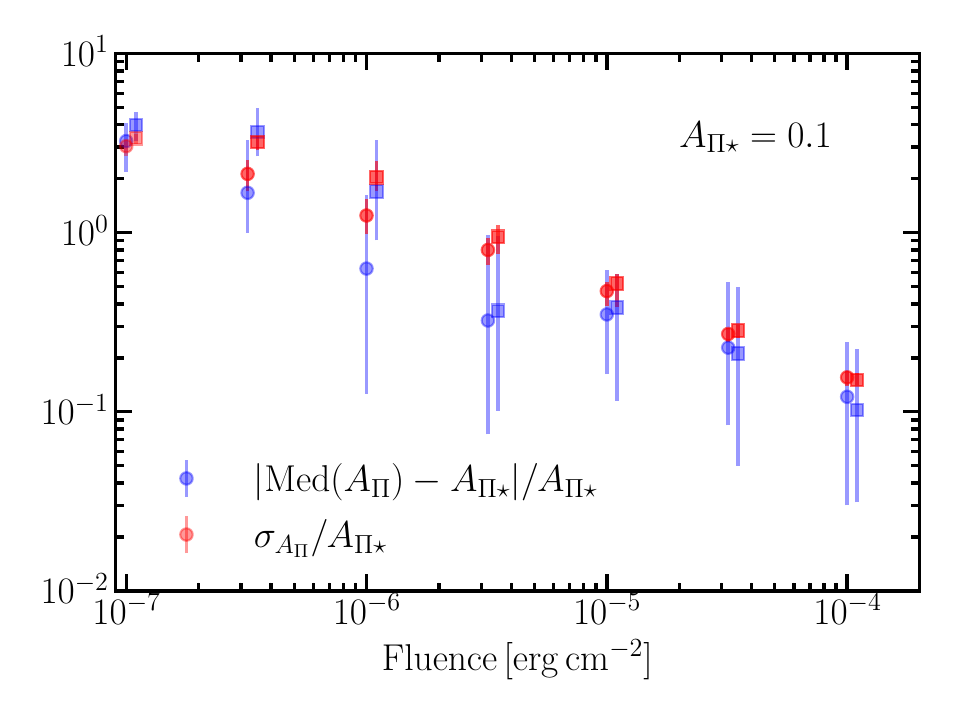}
    \includegraphics[width=0.48\textwidth]{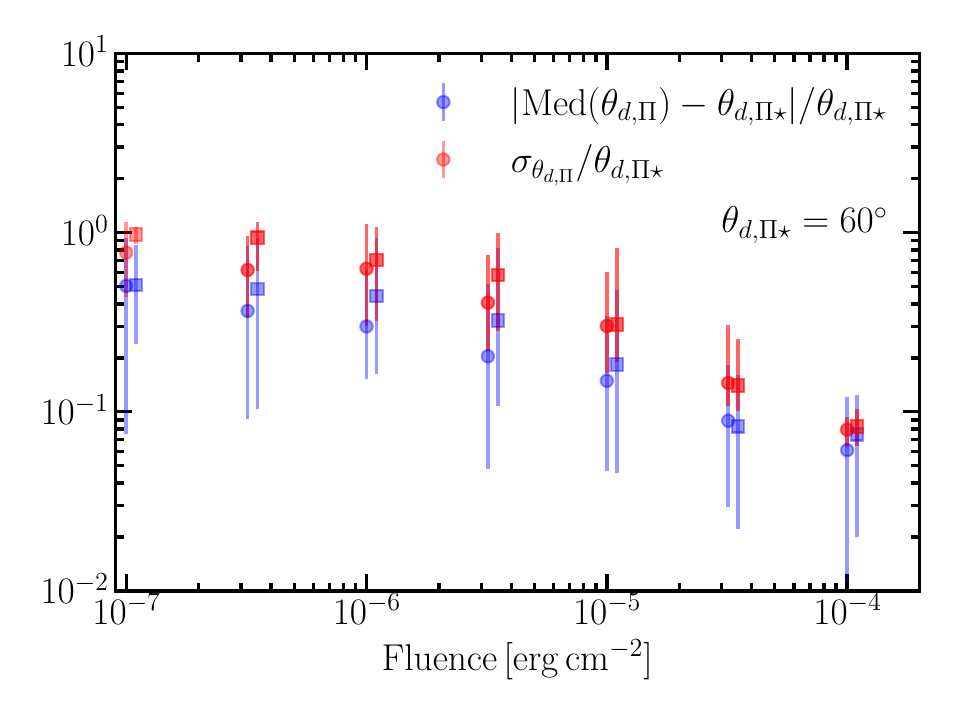}
    \caption{Accuracy and precision of the best-fit model parameters as a function of source fluence, obtained by performing 100 random realizations of the model prepared using the true values shown with a star in all panels. Accuracy (blue points) is shown by the difference between the true value and median (or best-fit) value of the posterior distribution of the model parameters obtained in each random realization. The precision (red points) is shown by the $1\sigma$ spread of the posterior distribution obtained in each random realization around the best-fit (or median) value. Data shown using circles only include source photons and that shown with squares (artificially shifted in fluence by a multiplicative factor of 1.1 to prevent overlapping) include background photons.
    }
    \label{fig:Pol-fit-error-Vs-fluence}    
\end{figure*}
%%%%%%% FIGURE %%%%%%%%%%%%%%%%%%%%%%%%%%%%%%%%%%%%%%%%%%%%%

Figure\,\ref{fig:A-and-PA-dist-with-bknd-photons} shows the posterior distributions for both $A$ and $\theta_\Pi$ from a single random realization (top-left panel) and from 300 random realizations (top-right panel), both prepared from the true solution with $A_{\Pi\star}=0.5$ and $\theta_{d,\Pi\star}=20^\circ$. The bottom panel shows the true modulation curve (red dashed) for the test source and the blue shaded region shows the $1\sigma$ uncertainty region around the best-fit solution. As shown below, the best-fit solution is only accurate to 10 per cent at a source fluence of $\mathcal{F}=10^{-5}\,{\rm erg\,cm^{-2}}$ for the chosen true values of the parameters. Hence the broad $1\sigma$ region. The pulse-integrated polarization \confirm{in the detector's energy range} for the chosen set of parameters is $\Pi_{B_{\rm ord}} = \confirm{55}$\,per cent, and therefore a value of $A_\Pi = 0.5$ corresponds to a pulse-integrated polarization of $\Pi = A\Pi_{B_{\rm ord}} = \confirm{27.5}$\,per cent.

The uncertainties in the two parameters as a function of source fluence, with a polarized test source prepared with and without background photons, is shown in Fig\,\ref{fig:Pol-fit-error-Vs-fluence} for two different sets of true parameter values. In general, the solution shows a significant spread in the accuracy and precision at all levels of fluence. This is caused by the reduced number of photons due to a sharp suppression of the Compton effective area of the instrument below $\sim30$\,keV. The top row of the figure shows the moderately polarized case and the bottom presents the weakly polarized case, both with different PAs. As the source polarization becomes weaker for a fixed level of fluence, the $1\sigma$ spread in the distribution of best-fit values grows. This means that a weakly polarized source is more likely to yield a greater variety of best-fit solutions for different random realizations. The relative accuracy of the best-fit values of course increases as $\propto\mathcal{F}^{-1/2}$ as the source fluence grows. It also increases for a higher level of $\Pi$, however, the absolute accuracy remains constant. For example, in the two cases shown in Fig\,\ref{fig:Pol-fit-error-Vs-fluence}, the more strongly polarized source (with pulse-integrated $\Pi = \confirm{27.5}$\, per cent) yields a relative accuracy of $\simeq8$\,per cent in determining the PD at $\mathcal{F}=10^{-5}\,{\rm erg\,cm^{-2}}$, while the weakly polarized source (with pulse-integrated $\Pi = \confirm{5.5}$\, per cent) yields a relative accuracy of $\simeq40$\, per cent, which is a factor of 5 larger corresponding to the factor of 5 increase in PD between the two cases. More importantly, the $1\sigma$ absolute accuracy in both cases is the same, which is $\simeq\confirm{2.2}\mathcal{F}_{-5}^{-1/2}$\, per cent for a pulse fluence of $\mathcal{F}=10^{-5}\mathcal{F}_{-5}\,{\rm erg\,cm^{-2}}$ as long as the source photons dominate over the background. Broadly similar conclusion can be drawn for the accuracy in determining the PA.

\begin{figure*}
    \centering
    \includegraphics[width=0.45\textwidth]{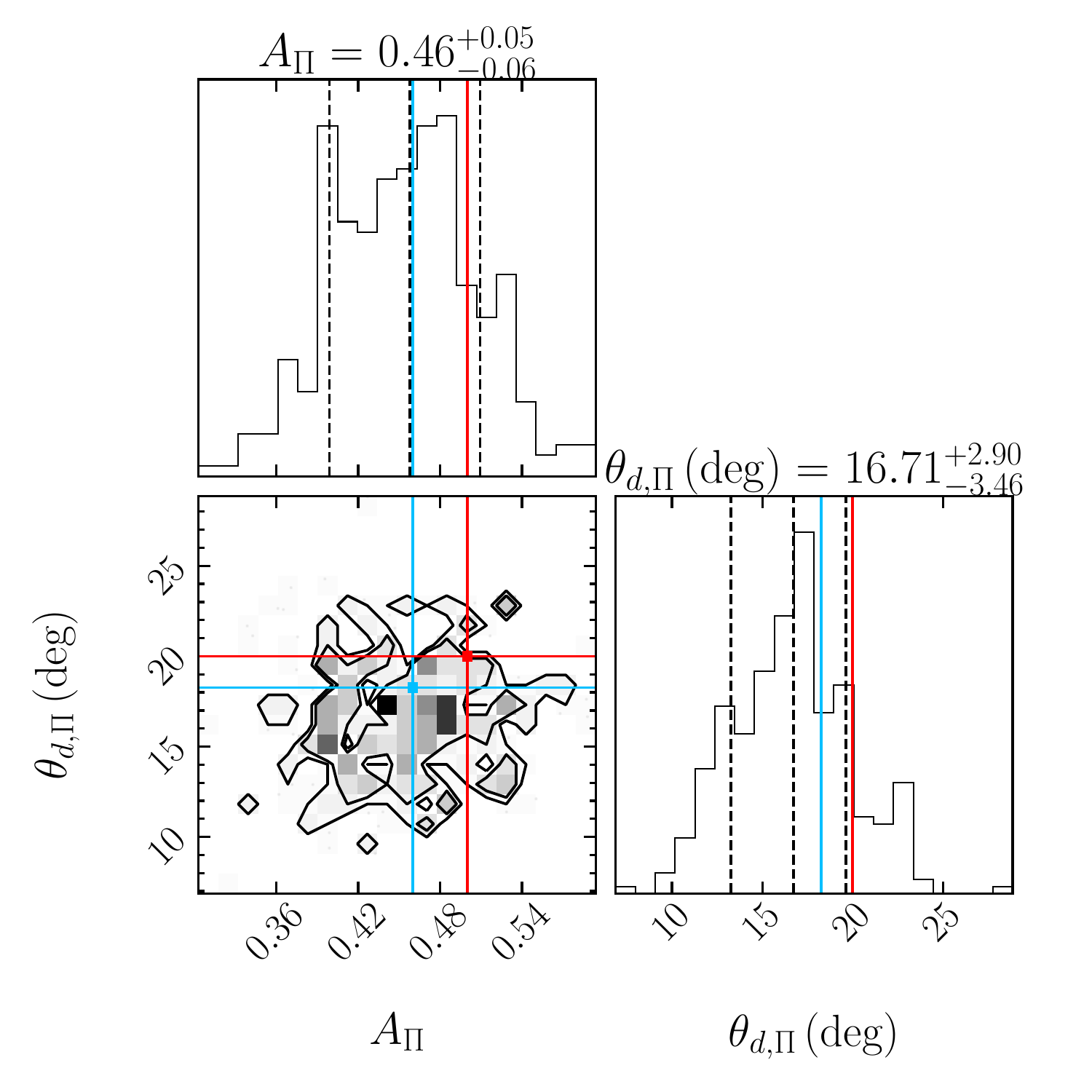}
    \includegraphics[width=0.48\textwidth]{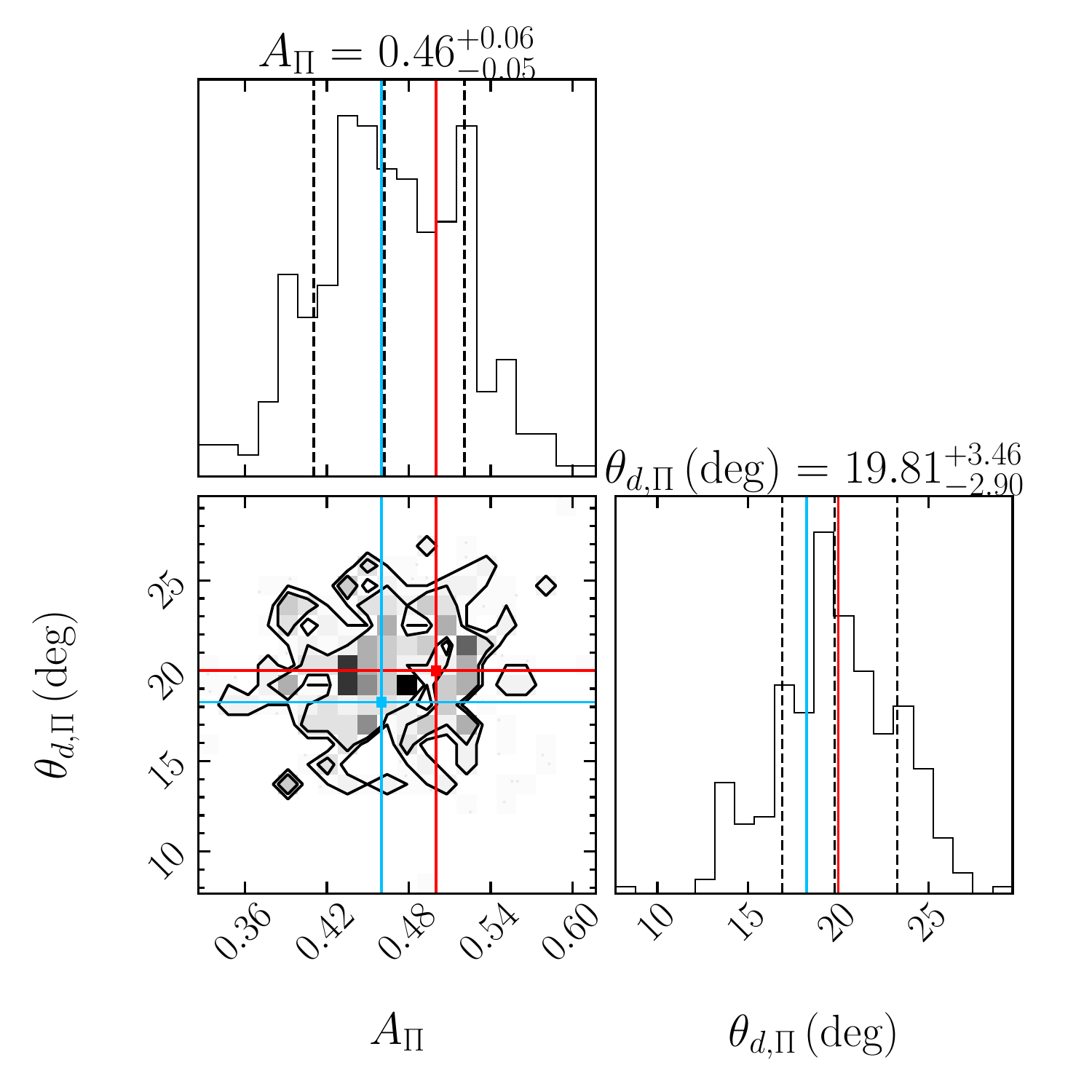}
    \caption{(\textbf{Left}) The posterior distributions for $A_\Pi$ and polarization angle $\theta_{d,\Pi}$ for source fluence $\mathcal{F}=10^{-5}\,{\rm erg\,cm}^{-2}$ with background photons, obtained from 300 random realizations now prepared using the best-fit solution (left panel of Fig.\,\ref{fig:A-and-PA-dist-with-bknd-photons}) from a single realization. 
    (\textbf{Right}) Inferred distributions of the true model parameters obtained from inverting (i.e. by removing the systematic bias) the distributions in the left panel. The red lines show the true model parameter values.
    }
    \label{fig:true-A-and-PA-dist-with-bknd-photons}
\end{figure*}

Next, we proceed to remove the systematic bias by preparing 300 random realizations using the best-fit solution obtained from a single realization. The left panel of Fig.\ref{fig:true-A-and-PA-dist-with-bknd-photons} shows the posterior distributions, where a clear bias is seen in the posterior distribution of $\theta_{d,\Pi}$. We correct for this bias in the right panel of the same figure and find that the peak of the distribution now is much closer to the true value of $\theta_{d,\Pi}$.

%%%%%%% FIGURE %%%%%%%%%%%%%%%%%%%%%%%%%%%%%%%%%%%%%%%%%%%%%
%%%%%%% FIGURE %%%%%%%%%%%%%%%%%%%%%%%%%%%%%%%%%%%%%%%%%%%%%

%%%%%%%%%%%%%%%%%%%%%%%%%%%%%%%%%%%%%%%%%%%%%%%%%%%%%%%
\section{Summary \& Discussion}\label{sec:discussion}
%%%%%%%%%%%%%%%%%%%%%%%%%%%%%%%%%%%%%%%%%%%%%%%%%%%%%%%
We have developed a novel method of model parameter estimation using maximum likelihood analysis of unbinned spectro-polarimetric observations of prompt GRB emission. The general practice is to first bin the data in both energy and time prior to fitting it with simple power-law models that are also binned. In addition, the fitting in time and energy is never performed simultaneously. Both methods, however, perform the fit by \textit{forward-folding} the theoretical model, i.e. convolving it with the detector response, and then fitting the observations in \textit{count space} rather than \textit{flux space}. The advantage of the unbinned method here is that it operates on the list of detected events (or photons) in energy and time and allows the fit to be performed simultaneously. Furthermore, there is no loss of information since the observations are in fact compared event by event with the model by calculating the likelihood of detecting any event at a given time and energy provided the theoretical model. In doing so it does require more computational effort in comparison to the binned method, however, the estimated model parameters can be constrained with much higher accuracy than possible with the binned method when the source fluence is low. The binned method is expected to perform equally well as the unbinned method when the source fluence is high. A future work will explore, among other issues, at what fluence level the transition to better accuracy by the unbinned over the binned method occurs. 

Using this new technique we have performed model fits to synthetic spectra and polarized emission over the duration of a single GRB pulse. We show that with POLAR-2's sensitivity we will be able to constrain the parameters in our model that control the jet dynamics to 1 per cent or better for GRBs with fluence $\mathcal{F}\gtrsim10^{-6}\,{\rm erg\,cm^{-2}}$. These include the radial distance over which the jet continuously radiates before switching off as well as the parameter that encodes whether the jet accelerates, decelerates, or simply coasts over that radial distance. 
This level of accuracy is afforded due to the high sensitivity of the theoretical model to the shape of the lightcurve and most importantly its well localized peak in time. The assumed constant background level for the assumed duration of $t_{\rm GRB}=10$\,s is high and even dominates the event rate at this source fluence. Since the background is assumed not to be time-varying, the fitting method is able constrain the dynamical parameters well given the highly time variable source signal. Of course, it is expected that the same level of accuracy will not be maintained if the background were to vary over similar timescales as the GRB, which is not realized in reality. The true background time-dependence will only be understood after POLAR-2 is deployed on the CSS. At that point it will be easily incorporated into our modelling that will yield more accurate results.

To constrain the spectral parameters well with the HPD instrument on board POLAR-2, a bright GRB with fluence $\mathcal{F}\gtrsim10^{-5}\,{\rm erg\,cm^{-2}}$ is needed. The culprit here is the high background that only begins to become subdominant for GRBs brighter than the quoted source fluence level. POLAR-2 will also feature a dedicated spectrometer (Broad-band Spectrometer Detector -- BSD), which is not used in this work but it offers better sensitivity over HPD. It is then expected that the spectrum will be better constrained even at a lower source fluence.

To reduce computational effort we do not perform a joint fit over the time-dependent spectrum and polarization in this work, although it is entirely possible to do that using our method. Once the dynamical and spectral parameters are constrained, they are fixed to the best-fit values when performing a polarimetric fit over the PD and PA. The latter is only a two parameter fit and involves a much reduced number of photons due to the suppression of the Compton effective area at lower energies. The LPD uses a gas detector to measure the polarization of 2-10 keV photons through the photoelectric effect. The BSD also has a certain ability to detect the polarization of high energy photons by means of the Compton effect. The cooperation of the three payloads will play an important role in studying the variation of polarization with energy for bright GRBs, which will help to distinguish between different competing models. The reader may refer to \citet{2024arXiv240714243Y}, \citet{2025arXiv251002016K} and \citet{Sun+25} for the polarization measurement capabilities of LPD and BSD.

With the greater sensitivity of POLAR-2 the PD of a modestly polarized GRB, having pulse-integrated $\Pi\gtrsim\confirm{27}$\,per cent, can be constrained to an accuracy of better than 10 per cent when the source fluence is $\mathcal{F}\gtrsim10^{-5}\,{\rm erg\,cm^{-2}}$. More generally, for a pulse fluence of $\mathcal{F}=10^{-5}\mathcal{F}_{-5}\,{\rm erg\,cm^{-2}}$ \confirm{and higher} the time-integrated PD can be constrained to an absolute accuracy ($1\,\sigma$) of about $\confirm{2.2}\mathcal{F}_{-5}^{\,-1/2}$ per cent as long as source photons dominate over the background. For example, \textit{Fermi}-GBM detects around $\sim\!100$ GRBs per year with total fluence $\mathcal{F}_{-5}>1$ \citep[see, e.g., Fig.\,1 of][]{Burns+23}, where most GRBs likely have complex emission episodes with multiple overlapping pulses. Even if a fraction of these bright GRBs that may have distinct single pulses and are seen by POLAR-2, since it has a smaller field-of-view when compared to GBM, it is likely to gather strong statistics on GRBs that are not strongly polarized. When comparing the field of views (FoVs) of the two instruments, with FoV$\;>8$\,sr for \textit{Fermi}-GBM and FoV$\;\approx6.28$\,sr (50\% of the sky) for POLAR-2, the latter may detect around $\sim78$ GRBs with a total fluence $\mathcal{F}_{-5}>1$. If it turns out that most or all of the POLAR-2 GRBs are polarized with pulse-integrated $\Pi\lesssim20$ per cent, the prompt emission model invoking a large-scale ordered magnetic field will be ruled out \citep{Gill+20}. On the other hand, a single high-significance detection of $\Pi\gtrsim50$ per cent will strongly favour such a model.

When looking at the prompt GRB polarization measurements \citep[see Fig.\,14 and Table\,1 of][]{Gill+21} from different instruments, those made by IKAROS-GAP \citep{Yonetoku+12} and AstroSAT-CZTI \citep{Gupta+24} show a very high PD with $\Pi\gtrsim40$ per cent for several GRBs, albeit with large $1\sigma$ errorbars. On the other hand, measurements made by POLAR \citep{Zhang+19a} favour a mean time-integrated PD of $\Pi\lesssim10$ per cent for a sample of five GRBs for which they measured the PD with high precision. Four of them had a fluence in excess of $10^{-5}\,{\rm erg\,cm^{-2}}$ in the ($10-1000$) keV energy range. With the sensitivity of POLAR-2 the PD of these kinds of GRBs and those detected by IKAROS-GAP and AstroSAT-CZTI would be constrained to a few per cent, which would finally settle the disparity between the results obtained by different instruments.

In this work we have only explored the scenario of a single-pulse GRB. Our theoretical model and fitting technique is also capable of testing scenarios with multiple overlapping pulses. While such cases would present significant model parameter degeneracies, if significant polarization ($\Pi\gtrsim30-40$ per cent) will still be inferred despite imperfect decomposition of the lightcurve into its constituent pulses then this will strongly favour synchrotron emission from an ordered magnetic field. Here we focus on analysing well sampled, single pulse, bright GRBs, which are the cleanest cases for polarimetric studies. \confirm{In addition to having only a single pulse, we also fixed $\Delta R/R_0=1$ and $m=0$ in our test case. This limits the emitting shell to be in the coasting phase, which is expected in the internal shocks scenario, but also to remain active only over a single dynamical time (i.e. doubling of its radius). In practice, we expect to obtain different values of these parameters in different GRBs, at least for $\Delta R/R_0$ if not for $m$, that would lead to more complex pulse shapes even in a single pulse. We also assumed the simplest, and more promising when it comes to polarization from an on-axis burst, case of an ordered B-field. If the B-field is shock-produced and axisymmetric around the local shock normal, and consequently around the LOS, then the expected polarization vanishes. As discussed in section\,\ref{sec:B-field-jet-structure}, in such cases, a significant net 
%non-vanishing 
polarization is only possible when the observer's LOS is near the edge of the jet core and the jet possesses strong angular structure (including sharp edges in some cases) within the visible region. All of these effects that include different B-field configurations, viewing angles, jet geometries, and off-axis position of the source in the detector plane will be the subject of a future work.}

The theoretical models of time and energy dependent prompt GRB polarization have matured to the point that they are able to predict the time-integrated \citep{Gill+20} and time-resolved \citep{Gill-Granot-21} evolution of PD and PA for both axisymmetric and non-axisymmetric \citep{Gill-Granot-24} jets. Having realized the tremendous potential of measuring prompt GRB polarization with high statistical significance, which may finally provide the long sought after answers to some of the fundamental questions in GRB physics, several high-energy polarimetry mission have been planned, with some slated for launch in the near future. Apart from the POLAR-2 mission, these include the LargE Area burst Polarimeter (LEAP; \citealt{McConnel+21}), COmpton Spectrometer and Imager (COSI; \citealt{Tomsick+24}), Daksha \citep{Bala+23}, and COMCUBE-S \citep{Franel+25}.

%%%%%%%%%%%%%%%%%%%%%%%%%%%%%%%%%%%%%%%%%%%%%%%%%%%%%%%
\section*{Acknowledgements}
%%%%%%%%%%%%%%%%%%%%%%%%%%%%%%%%%%%%%%%%%%%%%%%%%%%%%%%

This work was performed using funds from the PAPIIT-2023 (IA105823) grant, the Special Exchange Program A of Chinese Academy of Sciences (Grant No.~2H2025000112), the Special Program for Enhancing Original Innovation Capability of Chinese Academy of Sciences (Grant No.~292024000260), the National Natural Science Foundation of China (Grant No.~11961141013 and 12333007) and China's Space Origins Exploration Program. 

%%%%%%%%%%%%%%%%%%%%%%%%%%%%%%%%%%%%%%%%%%%%%%%%%%%%%%%
\section*{Data Availability}
%%%%%%%%%%%%%%%%%%%%%%%%%%%%%%%%%%%%%%%%%%%%%%%%%%%%%%%
 
No data was used in this work.



\bibliographystyle{mnras}
% ##### refs.bib tries to follow a particular style of listing citations. 
% ##### If you have your own .bib file, please keep it separate and add entries there. 
% ##### Multiple bib files can be listed below for use. 
\bibliography{refs}

%%%%%%%%%%%%%%%%%%%%%%%%%%%%%%%%%%%%%%%%%%%%%%%%%%

%%%%%%%%%%%%%%%%% APPENDICES %%%%%%%%%%%%%%%%%%%%%

\appendix
\newpage
\begin{onecolumn}
%%%%%%%%%%%%%%%%%%%%%%%%%%%%%%%%%%%%%%%%%%%%%%%%%%%%%%%%%%%%%%%%%%%
\section{Relating the locally measured PA to that given in J2000}
%%%%%%%%%%%%%%%%%%%%%%%%%%%%%%%%%%%%%%%%%%%%%%%%%%%%%%%%%%%%%%%%%%%
In the J2000 coordinate system (shown in Fig.\,\ref{fig:coordinate}), the unit vectors attached to the square detector are given by
\begin{eqnarray}
    \hat x_d &=& \cos\theta_{J,d}\,\cos\varphi_{J,d}\,\hat x_J + \cos\theta_{J,d}\,\sin\varphi_{J,d}\,\hat y_J - \sin\theta_{J,d}\,\hat z_J \\
    \hat y_d &=& -\cos\theta_{J,d}\,\sin\varphi_{J,d}\,\hat x_J + \cos\theta_{J,d}\,\cos\varphi_{J,d}\,\hat y_J - \sin\theta_{J,d}\,\hat z_J \\
    \hat z_d &=& \sin\theta_{J,d}\,\cos\varphi_{J,d}\,\hat x_J + \sin\theta_{J,d}\,\sin\varphi_{J,d}\,\hat y_J + \cos\theta_{J,d}\,\hat z_J\,.
\end{eqnarray}
The radial unit vector $\hat z_s$ pointing towards the GRB (source) can now be expressed in both the J2000 and detector coordinates
\begin{eqnarray}
    \hat z_s &=& \sin\theta_{J,s}\,\cos\varphi_{J,s}\,\hat x_J + \sin\theta_{J,s}\,\sin\varphi_{J,s}\,\hat y_J + \cos\theta_{J,s}\,\hat z_J \\
    &=& \sin\theta_{d,s}\,\cos\varphi_{d,s}\,\hat x_d + \sin\theta_{d,s}\,\sin\varphi_{d,s}\,\hat y_d + \cos\theta_{d,s}\,\hat z_d\,.
\end{eqnarray}
Since both $\hat z_d$ and $\hat z_s$ are well defined in the J2000 coordinate system, the off-instrument-axis angle of the source ($\theta_{d,s}$) in the detector plane is expressed as
\begin{equation}
    \cos\theta_{d,s} = \hat z_d \cdot \hat z_s = \sin\theta_{J,s}\sin\theta_{J,d}\cos(\varphi_{J,s}-\varphi_{J,d}) + \cos\theta_{J,s}\cos\theta_{J,d}\,.
\end{equation}
The plane of the sky is orthogonal to $\hat z_s$ and it is defined by two mutually orthogonal vectors
\begin{eqnarray}
    \hat x_s &=& \cos\theta_{J,s}\,\cos\varphi_{J,s}\,\hat x_J + \cos\theta_{J,s}\,\sin\varphi_{J,s}\,\hat y_J - \sin\theta_{J,s}\,\hat z_J \\
    \hat y_s &=& -\sin\varphi_{J,s}\,\hat x_J + \cos\varphi_{J,s}\,\hat y_J\,.
\end{eqnarray}
The two unit vectors, $\hat z_d$ and $\hat z_s$, also form a plane and the intersection of this plane with the plane of the sky results in another unit vector
\begin{equation}
    \hat x_0 = \cos\theta_{d,s}\,\cos\varphi_{d,s}\,\hat x_d + \cos\theta_{d,s}\,\sin\varphi_{d,s}\,\hat y_d - \sin\theta_{d,s}\,\hat z_d\,,
\end{equation}
which is on the plane of the sky. When the source is on-instrument-axis, i.e. $\theta_{d,s}=0$, we take $\varphi_{d,s}=0$ which gives $\hat x_0=\hat x_d$. The vector $\hat x_0$ is rotated away from $\hat x_s$ by an angle $\theta_0 = \arccos(\hat x_0 \cdot \hat x_s)$. 

The polarization unit vector is always contained in the plane of the sky, such that 
\begin{equation}
    \hat\Pi = \cos\theta_\Pi\,\hat x_s + \sin\theta_\Pi\,\hat y_s
\end{equation}
where $\theta_\Pi$ is the PA. Locally, using the distribution of the scattering azimuth in the detector, we measure the angle $\theta_{d,\Pi}=\theta_0 + \theta_\Pi$. By subtracting $\theta_0$ from this angle, where $\theta_0$ is well defined, we can express the PA using the convention shown in Fig.\,\ref{fig:coordinate}.

\end{onecolumn}
%%%%%%%%%%%%%%%%%%%%%%%%%%%%%%%%%%%%%%%%%%%%%%%%%%

% Don't change these lines
\bsp	% typesetting comment
\label{lastpage}
\end{document}